\documentclass[]{interact}

\usepackage{epstopdf}
\usepackage[caption=false]{subfig}
\usepackage{tabularx}
\usepackage{hyperref}

\theoremstyle{plain}

\theoremstyle{definition}

\theoremstyle{remark}

\begin{document}

\title{A Systematic Survey on Image Description Techniques for STEM Domains}

\author{
\name{Marco Cardia\textsuperscript{a}\thanks{CONTACT Marco Cardia. Author. Email: marco.cardia@di.unipi.it} 
and Letizia Angileri \textsuperscript{a}
and Marina Buzzi \textsuperscript{b}
and Giulio Galesi \textsuperscript{a}
and Barbara Leporini \textsuperscript{a,c}
}
\affil{\textsuperscript{a}University of Pisa, Pisa, Italy; \textsuperscript{b}CNR - IIT, Pisa, Italy; \textsuperscript{c}CNR - ISTI, Pisa,}
}

\maketitle

\begin{abstract}
The proliferation of visual data in Science, Technology, Engineering, and Mathematics (STEM) fields presents accessibility barrier for individuals with blindness or visual impairments.
While recent advances in Artificial Intelligence (AI) offer new opportunities to generate textual descriptions of STEM images, the research landscape is fragmented and its impact on real users remains limited.
This systematic survey examines 20 peer-reviewed studies on AI-based techniques for describing STEM visuals, with a specific focus on accessibility and human–computer interaction.
Following the PRISMA methodology and a ROBIS-based risk-of-bias assessment, the review analyzes (i) the types of STEM visuals targeted, (ii) the AI and machine learning architectures employed, (iii) the datasets and evaluation metrics adopted, and (iv) the interaction modalities through which descriptions are delivered. The analysis reveals a shift from static, one-shot alt text toward interactive and multimodal systems that integrate conversational interfaces, keyboard navigation, and audio or haptic feedback. However, critical challenges persist, including factual inaccuracies and hallucinations, the scarcity of accessibility-first datasets co-designed with blind and low-vision users, and a heavy reliance on automatic text-overlap metrics that poorly capture perceived usefulness and trust. The survey concludes by outlining key research directions for HCI, emphasizing user-controlled verbosity, explainable and verifiable AI pipelines, and the integration of accessible description tools into mainstream STEM authoring and learning environments.
\end{abstract}


\section{Introduction}
\label{sec:introduction}

In the domains of Science, Technology, Engineering, and Mathematics (STEM), visual representations are not merely illustrative aids; they are the primary medium for representing complex data, abstract concepts, and intricate relationships. Charts, graphs, diagrams, and equations form the basis of scientific discourse, enabling researchers to present evidence, educators to explain principles, and students to build mental models of complex systems \cite{mukhiddinov2021systematic}. The proliferation of digital media has further amplified this reliance, with scientific publications, online educational platforms, and collaborative research tools characterized by data visualizations. Consequently, the ability to interpret these visual artifacts has become a fundamental prerequisite for meaningful participation and advancement in any STEM field. These images are the language of science, and fluency is essential for comprehension and contribution \cite{brinn2022framework}. 

STEM images includes a wide range of categories, including scientific illustrations that depict anatomical structures or chemical processes, mathematical representations such as graphs and equations used to support quantitative reasoning, and computer science diagrams representing algorithms, system architectures, or data structures. While these images enrich scientific communication, they simultaneously create accessibility barriers for the millions of individuals globally who are blind or have low vision (BLV) \cite{shahira2021towards}.
When critical information is encoded exclusively in images, BLV learners and professionals cannot independently access the evidence on which scientific claims rely. This exclusion is not merely an inconvenience: it introduces a form of inequity, restricting educational opportunities, professional inclusion, and, even more profoundly, the ability to participate in the scientific method itself, which depends on the independent interpretation and evaluation of data \cite{mukhiddinov2021systematic, poggianti2025immersive}. Traditional accessibility resources such as tactile graphics can assist in specific cases, but they are costly to produce, difficult to scale to the vast and rapidly growing corpus of digital STEM materials, and insufficient for interactive or dynamic content.

To address this inequity, image description (or captioning), which consists of producing textual explanations capturing the content and meaning of an image, serves as a critical mechanism for inclusion. Within STEM, descriptions play a dual role: they facilitate access for BLV users and support the comprehension of complex visual material in broader educational contexts. Traditionally, accessibility has relied on manual curation. While human-authored captioning ensures accuracy, it is resource-intensive, demands domain expertise, and cannot keep pace with the exponential growth of visual scientific content. Crowdsourced or non-expert annotations, on the other hand, often lack the technical precision required for STEM contexts.

This scalability bottleneck has incresed the interest in automated approaches. Rapid advancements in Artificial Intelligence (AI), specifically in Computer Vision (CV), Natural Language Processing (NLP), and Natural Language Generation (NLG), offer transformative potential. AI-driven systems can automatically analyze the content of STEM images and generate rich textual descriptions that can be rendered by assistive technologies, thereby unlocking vast repositories of scientific knowledge and supporting greater independence for BLV users. Preliminary tools, such as AlternAtIve \cite{pedemonte2025improving}, demonstrate early progress in producing domain-aware alt-text. However, other evaluations indicate that mainstream generative models still struggle with technical accuracy \cite{buzzi2024isgenerative}. Recent research has explored the role of AI in inclusive education, highlighting how generative AI can improve accessibility and personalize learning for students with disabilities or special needs \cite{bernardi2016automatic, melo2025impact}. However, only a limited subset of this work directly addresses the specific challenge of generating accessible descriptions for STEM images, which require levels of precision and domain knowledge not typically encountered in general captioning tasks.

Yet, the automation of STEM captioning presents a unique set of challenges that distinguishes it significantly from general-purpose image description. Unlike natural images, where identifying objects and basic spatial relationships may suffice, STEM visuals require a multi-layered understanding. This includes recognizing graphical elements (bars, lines, points), extracting precise data values, identifying statistical relationships (correlations, trends, causality), parsing specialized syntax (mathematical notation), and interpreting the overall semantic meaning within the scientific context \cite{singh2024figura11y, leotta2024evaluating}. Crucially, the cost of error in this domain is high. An AI system that "hallucinates" or generates a factually incorrect description of a chart does not merely mislabel an image, it distorts scientific evidence. Such inaccuracies can mislead students, compromise research, or propagate incorrect conclusions in fields where precision is foundational \cite{buzzi2024isgenerative}.

Therefore, the challenge extends beyond technical accuracy and must be addressed from a Human–Computer Interaction (HCI) perspective. It is both a technical and an interactional problem involving usability, trust, and cognitive load. AI-generated descriptions are embedded in workflows that span screen readers, authoring tools, educational platforms, and conversational agents. As such, they fundamentally shape how BLV users construct mental models, navigate complex information, and engage in scientific reasoning. Effective solutions must synthesize robust CV and NLG techniques with user-centered interaction paradigms to ensure that descriptions are not only correct but also navigable, interpretable, and aligned with users’ cognitive strategies.

Beyond their essential role in accessibility, alternative text and structured descriptions of STEM visuals contribute to a broader computational ecosystem. They support indexing, retrieval, and automated reasoning across digital libraries and scholarly infrastructures. High-quality descriptions enable search engines and multimodal AI systems to classify and interpret visual scientific content that would otherwise remain opaque to text-based pipelines. Moreover, alt-text serves as a critical supervisory signal for training vision–language models and image-generation systems, which require semantically grounded annotations to represent STEM concepts faithfully. The absence of accurate domain-specific descriptions therefore restricts both accessibility for BLV users and the development of advanced multimodal AI systems. From an HCI perspective, richer and more structured descriptions also unlock new forms of interaction, including conversational exploration of figures, semantic navigation of visual elements, and user-driven querying of scientific diagrams.

Given the fragmentation of current research, which is often siloed into specific tools, isolated datasets, or narrow case studies, there is a pressing need for a comprehensive synthesis of the methods, findings, and limitations. This paper addresses that gap by providing a structured and critical survey of AI-powered techniques for generating descriptions of STEM images. To ensure transparency, reproducibility, and methodological rigor, the review adheres to the PRISMA (Preferred Reporting Items for Systematic Reviews and Meta-Analyses) framework for study identification, screening, and selection, and employs ROBIS (Risk of Bias in Systematic Reviews) to assess potential sources of bias in the included studies \cite{page2021prisma,whiting2016robis}.

The objective of this work is to systematically map, categorize, and analyze the state of the art, describing significant research gaps, and propose directions for future investigation. To achieve this, the survey is guided by the following Research Questions (RQs):
\begin{enumerate}
    \item [\textbf{RQ1}:] What types of visual STEM content are being targeted for accessibility in current research? 
    \item [\textbf{RQ2}:] What technological approaches, specifically AI and Machine Learning (ML) models, are used to generate accessible descriptions? 
    \item [\textbf{RQ3}:] How is the quality of generated content evaluated, and what datasets, benchmarks, and evaluation metrics are used? 
    \item [\textbf{RQ4}:] What forms of description, interaction, and accessibility tools are available for users with visual impairments?
    \item [\textbf{RQ5}:] What are the current limitations and future research opportunities in making STEM visual information accessible through AI? 
\end{enumerate}

The remainder of this paper is structured as follows.
Section \ref{sec:methodology} details the systematic methodology employed for study selection and data analysis, adhering to the PRISMA framework.
Section \ref{sec:results} presents the core results of the survey, organized along the research questions.
Section \ref{sec:discussion} discusses the major gaps and challenges identified from the results.
Finally, Section \ref{sec:conclusion} concludes the paper with a summary of its key contributions.

\section{Methodology}
\label{sec:methodology}
To ensure a comprehensive, transparent, and replicable review, this study employed a systematic search strategy based on the Preferred Reporting Items for Systematic reviews and Meta-Analyses (PRISMA) methodology \cite{page2021prisma}. The process was divided into several distinct stages, from defining the research questions to the final data synthesis. The PRISMA methodology includes five phases: (i) Research questions specifications, (ii) definition of the research criteria, (iii) identification of the studies, (iv) screening, (v) results. Each of these stages are detailed in the following sections. 

\subsection{Research Questions (RQs)}
\label{subsec:research_questions}
The study was guided by the five RQs outlined in the introduction. Each question was designed to examine a distinct aspect of the research landscape: 

\begin{itemize}
    \item \textbf{RQ1}: \textit{What types of visual STEM content are being targeted for accessibility in current research?} It aims to map the focus of the research community. This is critical for identifying which types of scientific visuals have received significant attention and which remain under-explored, revealing biases and opportunities in the field.
    \item \textbf{RQ2}: \textit{What technological approaches, specifically AI and Machine Learning (ML) models, are used to generate accessible descriptions?} It seeks to create a taxonomy of the underlying technologies. Understanding the architectural choices, from early machine learning models to modern deep learning and generative AI, is essential for tracing the field's technical evolution and understanding the trade-offs involved. 
    \item \textbf{RQ3}: \textit{How is the quality of generated content evaluated, and what datasets, benchmarks, and evaluation metrics are used?} It critically examines the methodologies used to validate these systems. Assessing the reliance on automated metrics versus human-centered evaluation, and the availability of robust datasets, is crucial for gauging the true maturity and real-world applicability of the reported solutions. 
    \item \textbf{RQ4}: \textit{What forms of description, interaction, and accessibility tools are available for users with visual impairments?} It investigates the nature of the output provided to the end-user. This moves beyond the generation model to the user experience, distinguishing between static text dumps and more sophisticated, user-driven interaction paradigms. This question also examines the alignment of current AI-generated descriptions with standards for accessibility. 
    \item \textbf{RQ5}: \textit{What are the current limitations and future research opportunities in making STEM visual information accessible through AI?} 
\end{itemize}

\subsection{Search Strategy}
\label{subsec:search_strategy}
We designed a search strategy to identify relevant primary studies from major academic databases. We selected the following five digital libraries for their extensive coverage of computer science, engineering, accessibility, and biomedical literature: IEEE Xplore, ACM Digital Library, Scopus, Web of Science (WoS), and PubMed. The choice to include multiple academic databases in our search strategy was driven by the goal of ensuring broad, multidisciplinary, and methodologically robust coverage of research on AI-driven techniques for generating accessible descriptions of STEM images. Scopus was included for its broad international scope and its extensive indexing across computer science, engineering, education, and cognitive sciences, all highly relevant to the accessibility of visual STEM content. IEEE Xplore and ACM Digital Library were selected to ensure thorough coverage of technical contributions in computer vision, machine learning, and human-computer interaction, which form the methodological backbone of current description-generation systems. PubMed was incorporated to capture studies at the intersection of accessibility, cognition, and assistive technology, offering insight into user needs and evaluation perspectives. Finally, WoS complemented these sources by providing access to high-quality, multidisciplinary research. Collectively, these databases ensured a diverse and representative corpus suited to analyzing the technological, educational, and accessibility-oriented dimensions of this field. 

\begin{figure}
    \centering
    \includegraphics[width=1\linewidth]{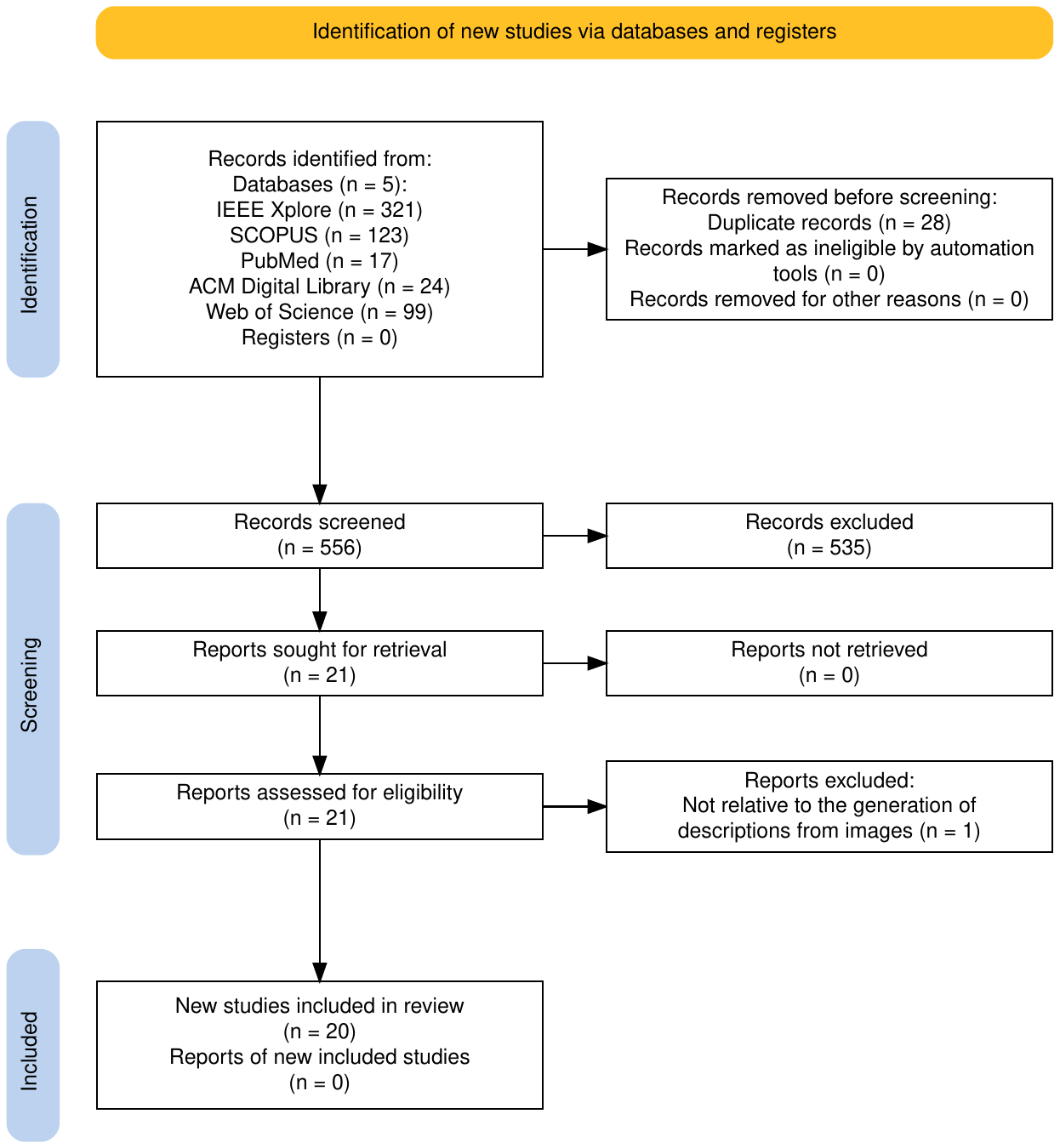}
    \caption{Flowchart showing the PRISMA selection process}
    \label{fig:prisma_flowchart}
\end{figure}

Figure \ref{fig:prisma_flowchart} shows the phases 3, 4, and 5 of the PRISMA methodology, reporting the number of papers included and excluded together with the reason. 

Preliminary exploratory searches on AI-generated descriptions of STEM images revealed that relevant studies are often distributed across diverse research communities, such as computer vision, accessibility, educational technology, and multimodal AI, rather than clustered under a single unified terminology. As a result, restricting the search to narrow or domain-specific keywords would have overlooked a substantial portion of the literature addressing accessibility, chart understanding, or STEM-specific visual interpretation. To ensure comprehensive coverage of research relevant to the automatic generation of accessible descriptions for STEM content, we adopted a broad and systematic keyword strategy. 
We constructed a Boolean search query by combining five concept groups with the AND operator. Each group contained a set of synonymous or related keywords connected by the OR operator. The use of wildcards (*) ensured the inclusion of lexical variations of search terms. The final Search Query was:

\texttt{("text generation" OR "language description*" OR "image description*" OR "educational description*" OR "accessible description*" OR "alternative text" OR "image label*" OR caption* OR "alt text")  
AND (image OR graph OR diagram OR chart OR cycle OR flowchart OR automata OR "Finite state machine*") 
AND (blind OR "visually impaired" OR accessib* OR inclusion OR "vision loss")  
AND (stem OR science OR technolog* OR math* OR engineering OR education OR geometr* OR physics OR chemistry OR biology OR ICT)  
AND (AI OR "Artificial Intelligence" OR "Machine Learning" OR ML OR "Generative AI" OR DL OR "Deep Learning" OR "Neural Network" OR NN OR ANN OR "Large Language Model")}

We identified the initial pool of studies by searching titles, abstracts, and keywords across the five selected electronic databases, each chosen for its relevance to computer vision, artificial intelligence, accessibility research, and STEM education. The search was conducted in June 2025. This process yielded a total of 584 records: Scopus = 123, IEEE Xplore = 321, Web of Science = 99, ACM Digital Library = 24, and PubMed = 17. Because many studies appeared in multiple databases, we removed 28 duplicate entries during the identification phase. After deduplication, 556 records remained for title and abstract screening. 

\subsection{Study Selection Protocol}
\label{subsec:study_selection_protocol}
As indicated in the PRISMA guidelines a rigorous, multi-stage screening process was used to select primary studies based on predefined inclusion and exclusion criteria. In line with the methodological characteristics that define research on AI-based generation of accessible STEM image descriptions, the inclusion criteria (IC) are defined as follows: 

\begin{itemize}
    \item \textbf{IC1}. The primary contribution of the paper must involve an automated or semi-automated method using AI or ML for generating textual descriptions from image content. 
    \item \textbf{IC2}. The target images must be explicitly identified as STEM-related (e.g., charts, graphs, diagrams, mathematical equations, scientific figures). 
    \item \textbf{IC3}. The primary motivation, application, or evaluation context must be accessibility for individuals who are blind or have low vision. 
\end{itemize}

The exclusion criteria (EC) are defined as follows: 

\begin{itemize}
    \item \textbf{EC1}. Studies focused on general-purpose image captioning of natural scenes (e.g., photographs of people, animals, landscapes) without a specific focus on STEM content or accessibility for BLV users. 
    \item \textbf{EC2}. Studies focused solely on manual authoring guidelines, non-AI accessibility techniques (e.g., tactile graphics generation without an AI-driven descriptive component), or usability evaluations of existing tools without proposing a new generation method. 
    \item \textbf{EC3}. Papers not written in English, not peer-reviewed (e.g., workshop papers, dissertations, most pre-prints), or not available in full-text format. 
\end{itemize}

\subsection{Execution and Reporting}
\label{subsec:execution_and_reporting}
The review process was executed systematically. First, the search query was run across all selected databases, and the results were aggregated. Duplicate entries were programmatically and manually removed. Next, all researchers independently screened the titles and abstracts of the remaining articles against the selection criteria. The full texts of the articles that passed the initial screening were then retrieved and reviewed in detail by two researchers to make the final inclusion decision. This process and its outcomes are summarized in the PRISMA flow diagram, shown in Figure \ref{fig:prisma_flowchart}. 

\subsection{Data Extraction and Synthesis}
\label{subsec:data_extraction_and_synthesis}
For each study included in the final corpus, a structured data extraction form was used to collect relevant information. The extracted data points were organized according to a set of predefined dimensions designed to answer the research questions: 
\begin{itemize}
    \item \textbf{Target Visual Content}: The specific type(s) of STEM images addressed (e.g., bar chart, flowchart). 
    \item \textbf{AI/ML Architecture}: The core algorithms and model architectures employed (e.g., CNN, Transformer, LLM). 
    \item \textbf{Description Modality \& Interaction Style}: The nature of the generated output (e.g., static alt text, interactive Q\&A). 
    \item \textbf{Datasets \& Benchmarks}: The corpora used for training and/or evaluation. 
    \item \textbf{Evaluation Paradigm}: The methods used to assess quality, including automated metrics and details of any user involvement. 
\end{itemize}

\subsection{Risk of Bias Assessment}
\label{subsec:risk_of_bias_assessment}
To assess the methodological quality of the included studies, the ROBIS (Risk of Bias in Systematic Reviews) tool was considered for its applicability in evaluating potential biases in the research process and reporting of the primary studies \cite{whiting2016robis}. The extracted data was then qualitatively synthesized to identify patterns, trends, and gaps across the literature, forming the basis of the results presented in the following section. 

Since our survey addresses AI-driven accessibility of STEM visuals rather than the healthcare-focused scenarios for which ROBIS was originally designed, we introduced minor methodological adjustments, explicitly indicated by the tool’s guidelines, to better align the assessment with the characteristics of this research domain \cite{whiting2016robis}.  

The first phase of the ROBIS tool focuses on assessing relevance. To structure this step, we applied the Population, Intervention, Comparator, Outcome (PICO) framework, adapting it to the specific characteristics of our research domain as follows: 
\begin{itemize}
    \item Population: studies involving individuals, such as students, educators, researchers, or people who are blind or have low vision, who engage with STEM visual content in educational, scientific, or accessibility-related contexts. 
    \item Intervention: AI-based methods for generating textual descriptions of STEM images, including charts, diagrams, equations, or other scientific figures. 
    \item Comparator: not applicable in our context. In line with ROBIS recommendations, this component was substituted with a thematic categorization of the included studies according to the five analytical dimensions corresponding to our RQs. 
    \item Outcomes: qualitative and quantitative evidence related to the generation, evaluation, and accessibility of AI-produced descriptions, synthesized according to the five RQs. 
\end{itemize}

This initial stage showed a strong consistency between the aims of the survey and the research questions that were defined. 

The second phase of the ROBIS tool focuses on identifying potential concerns in the review process. To conduct the assessment, we addressed the signaling questions within the four ROBIS domains. Answers were classified as Yes (Y), Probably Yes (PY), Probably No (PN), No (N), or Not Applicable (NA). Our responses and the overall rating are reported below. 

\textbf{[DOMAIN 1]} - \textit{Study Eligibility Criteria}

1.1 Did the review adhere to pre-defined objectives and eligibility criteria? Y 

1.2 Were the eligibility criteria appropriate for the review question? Y 

1.3 Were eligibility criteria unambiguous? Y 

1.4 Were any restrictions in eligibility criteria based on study characteristics appropriate (e.g. date, sample, size, study quality, outcomes measured)? Y 

1.5 Were any restrictions in eligibility criteria based on sources of information appropriate (e.g. publication status or format, language, availability of data)? PY

\textbf{[DOMAIN 2]} - \textit{Identification and Selection of Studies }

2.1 Did the search include an appropriate range of database/electronic sources for published and unpublished reports? Y 

2.2 Were methods additional to database searching used to identify relevant reports?PN 

2.3 Were the terms and structure of the search strategy likely to retrieve as many eligible studies as possible? Y 

2.4 Were restrictions based on date, publication format, or language appropriate? PY 

2.5 Were efforts made to minimize error in selection of studies? Y 

\textbf{[DOMAIN 3]} - \textit{Data Collection and Study Appraisal}

3.1 Were efforts made to minimise error in data collection? PY 

3.2 Were sufficient study characteristics available for both review authors and readers to be able to interpret the results? Y 

3.3 Were all relevant study results collected for use in the synthesis? PY 

3.4 Was risk of bias (or methodological quality) formally assessed using appropriate criteria? Y 

3.5 Were efforts made to minimise error in risk of bias assessment? Y 

\textbf{[DOMAIN 4]} - \textit{Synthesis and Findings}
4.1 Did the synthesis include all studies that it should? Y 

4.2 Were all pre-defined analyses reported or departures explained? Y 

4.3 Was the synthesis appropriate given the nature and similarity in the research questions, study designs and outcomes across included studies? Y 

4.4 Was between-study variation (heterogeneity) minimal or addressed in the synthesis? NA 

4.5 Were the findings robust, e.g. as demonstrated through funnel plot or sensitivity analyses? NA 

4.6 Were biases in primary studies minimal or addressed in the synthesis? NA 

Within Domain 2, the PN judgement is associated with signaling Question 2.2, related to the adoption of search strategies other than database queries. In this survey, we deliberately excluded secondary sources (e.g., grey literature and handsearching) to preserve replicability and to keep the corpus more focused. 

The third stage of the ROBIS tool focuses on synthesizing the concerns identified in Phase 2 in order to determine the overall risk of bias. Since the last signaling question in this phase, which addresses the importance of statistical significance, does not apply to our work (given the classificatory and non-comparative nature of our analysis), we only responded to questions A, and B: 

A. Did the interpretation of findings address all of the concerns identified in Domains 1 to 4? PY 

B. Was the relevance of identified studies to the review's research question appropriately considered? Y 

At this final stage, we concluded that the survey findings were in line with the concerns highlighted in Domains 1–4 and that the studies included were appropriate and pertinent to the objective of our review. 

\begin{figure}
    \centering
    \includegraphics[width=1\linewidth]{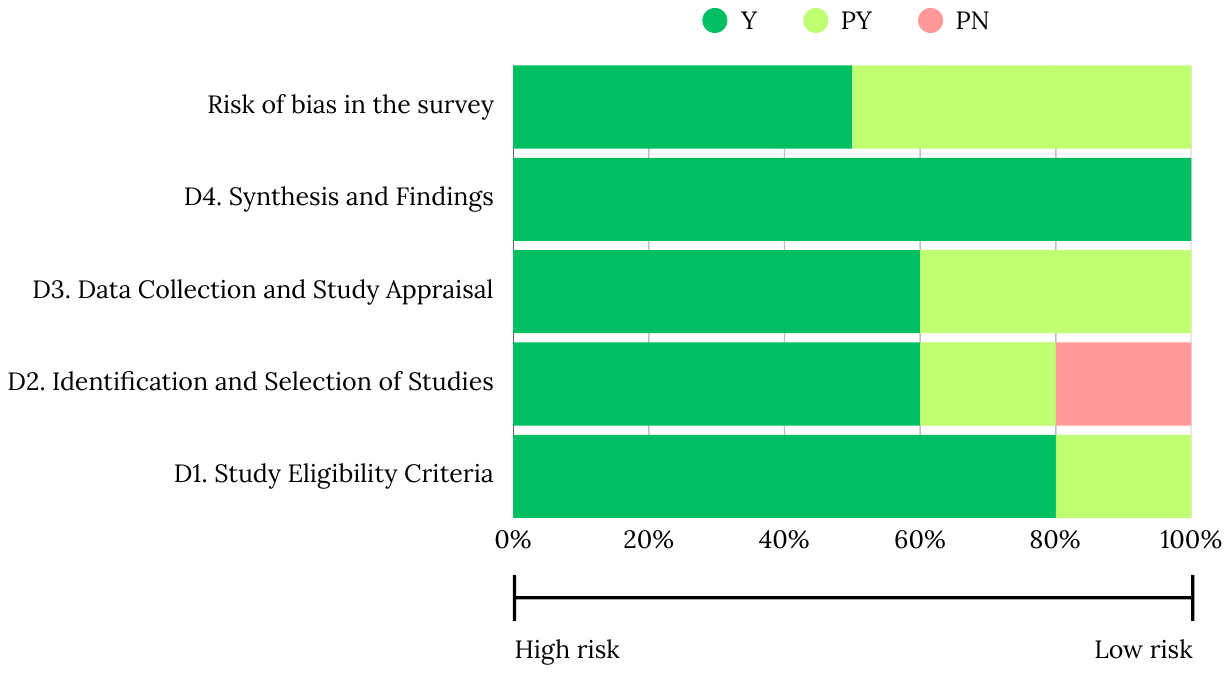}
    \caption{Overall ROBIS risk of bias evaluation. Normalized scores for each Phase 2 domain and the Phase 3 overall judgement; higher scores indicate a lower risk of bias.}
    \label{fig:robis}
\end{figure}

To obtain a visual overview of the final outcome, we assigned a numerical weight to each signaling question according to the following scale: Y = 20\%, PY = 15\%, PN = 5\%, N = 0\%. Using this scheme, we computed a composite score for each Domain in Phase 2 and for the signaling questions in Phase 3, and then generated the chart reported in Figure \ref{fig:robis}. Considering both the individual ratings and the aggregated scores, the overall risk of bias in our systematic survey was judged to be low. 

\subsection{Analysis of the results}
\label{subsec:analysis_of_the_results}
\begin{figure}
    \centering
    \includegraphics[width=1\linewidth]{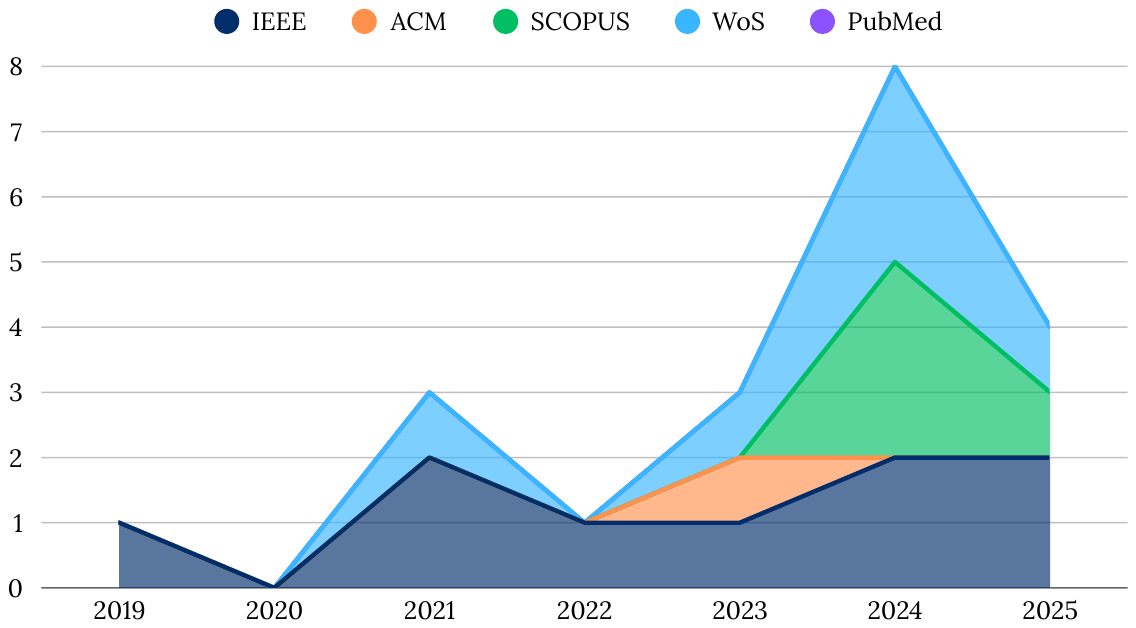}
    \caption{Number of publications per year for each database.}
    \label{fig:trends}
\end{figure}

Between January 2019 and June 2025, scientific production involving the generation of accessible alternative text for STEM images, including graphs, diagrams, and mathematical formulas, showed a marked increase. After a relatively modest publication rate until 2022, the field experienced significant expansion starting in 2023. There was a notable expansion in publication output, peaking in 2024 when eight contributions appeared in leading indexing platforms including IEEE, WoS, and Scopus. 
Although only the first half of 2025 is considered, the four publications already recorded suggest that the year is poised to at least match, if not exceed, the positive trend of the previous year. This trend reflects the recognition of the importance and urgency of accessibility research in STEM communication. 
Figure \ref{fig:trends} summarizes the publication trends over the years for each database. PubMed does not appear in the plot, as our survey did not include any studies indexed in this database.

\section{Results}
\label{sec:results}
In this section, we present the insights derived from the qualitative and quantitative analysis of the selected studies. We begin with an overview of the examined literature, followed by dedicated subsections addressing each of the five previously defined research questions. This structure allows for a clear and comprehensive synthesis of how the reviewed works contribute to the field. 

\subsection{Types of visual STEM content (RQ1)}
\label{subsec:rq1}
The focus of current accessibility research concerning visual STEM content is predominantly on data visualizations (charts and plots), followed by scientific and technical diagrams, and mathematical expressions. Mapping this focus helps reveal which areas are receiving significant attention and which remain under-explored. In this section, we address and describe the different types of visual STEM content being targeted in the literature. 

\subsubsection{Data visualization (Charts and plots)}
\label{subsubsec:data_visualization}
Data visualizations are fundamental tools for communication, enabling the efficient transmission of complex information through structured, interpretable representations. However, these images often present significant accessibility challenges for individuals who are blind or have low vision (BLV).  These kinds of figures are the most extensively studied type of STEM content in the literature.  

Research and data generation efforts primarily focus on fundamental and common chart types for accessibility improvements. These are generally considered easier to analyze: 

\begin{itemize}
    \item Bar Charts: These are a prominent focus in accessible visualization research. Efforts include methods for data extraction, chart summarization, and enabling rich interaction \cite{shahira2021towards,hoppe2021evaluation,farahani2023automatic,yan2025chart}. 
    \item Line Charts: Crucial for conveying trends over time, line charts are central to research, appearing in nearly all major datasets. Research addresses line charts, especially in the context of data visualization captioning models \cite{hoppe2021evaluation}. This includes time-series line charts and the challenges posed by complex trends or multi-lined charts \cite{hoppe2021evaluation,farahani2023automatic}. 
    \item Scatter Plots: These graphics, which show the correlation between two variables, are also commonly targeted for data extraction and accessible description. The goal is to accurately extract data and relationships from these plots \cite{shahira2021towards, farahani2023automatic,yan2025chart,nylund2025matplotalt}. 
    \item Pie Charts: These charts, used for proportional representation, are also included in foundational datasets for visual reasoning \cite{shahira2021towards,hoppe2021evaluation,farahani2023automatic,yan2025chart,nylund2025matplotalt}. 
\end{itemize}

While basic charts are well-represented, accessibility challenges persist for more intricate graphics, revealing areas that are relatively under-explored: 

\begin{itemize}
    \item Grouped, Multi-Series, and Panel Charts: These charts contain many data points and complex relationships between subplots \cite{shahira2021towards,yan2025chart}]. Researchers acknowledge that descriptions for these types of charts are harder to generate effectively compared to simpler charts \cite{yan2025chart}. Some systems, like Alt4Blind, specifically include examples of Multivariate and Panel charts \cite{yan2025chart,moured2024alt4blind}. 
    \item Choropleth Maps: These specialized geographical visualizations, which use colour to represent data density, were included in the stimulus set used to evaluate the VizAbility system \cite{gorniak2024vizability}. 
    \item Data Dashboards: Tools are beginning to explore accessibility for multiple-coordinated charts or dashboards, which present collections of dynamic data visualizations \cite{gorniak2024vizability}. 
    \item Scientific and Real-World Plots: Datasets like PlotQA specifically focus on reasoning over plots created from real-world scientific data, attempting to bridge the gap between synthetic data and actual data complexity. SciCap focuses on figures extracted from academic papers mostly composed by charts \cite{shahira2021towards,hsu2021scicap,yang2024scicap+,farahani2023automatic,yan2025chart}. 
    \item Other Chart Types: Accessibility efforts also extend to area charts \cite{yan2025chart,nylund2025matplotalt}, and heatmaps \cite{nylund2025matplotalt}. 
\end{itemize}

\subsubsection{General STEM Images and Diagrams}
\label{subsubsec:general_stem_images_and_diagrams}
Beyond data charts, research targets other critical visual components in scientific and technical literature, particularly complex images where details are crucial. Studies investigate the effectiveness of generating descriptive alt-text for various natural science illustrations and schemes: 
\begin{itemize}
    \item Biological illustrations. Examples include the Dolphin Anatomy illustration \cite{pedemonte2025improving}. Another figure example used in accessibility evaluation is the Photosynthesis schema \cite{leotta2024evaluating,pedemonte2025improving}. 
    \item Physical concepts. This includes figures illustrating concepts like Lunar Eclipse and the Parabolic Trajectory, often represented by a volcano image in experiments \cite{leotta2024evaluating,pedemonte2025improving}. 
    \item Chemical diagrams. Research has considered haptic representations for chemical formulas and textual descriptions for chemical diagrams \cite{balestrucci2024educational}. An image of the Krebs cycle was used to evaluate AI captioning tools for moderate difficulty content \cite{buzzi2024isgenerative}. 
    \item Figures from scholarly papers. There are challenges for accurately describing scientific figures within research papers, driving the development of specialized datasets focusing on captions for scientific figures \cite{shahira2021towards,farahani2023automatic,yan2025chart}. However, most of these images are charts \cite{shahira2021towards,farahani2023automatic,nylund2025matplotalt}. 
    \item Engineering diagrams. Examples include figures illustrating types of telescopes and types of bridges \cite{leotta2024evaluating,pedemonte2025improving}. 
    \item Computer science diagrams. Key content addressed includes graphical structures central to computer science education and research: 
    \begin{itemize}
        \item Finite State Diagrams (FSA). Dialogue systems have been developed and evaluated specifically as an alternative method for conveying information about FSA, which are typically represented through state diagrams or tables \cite{balestrucci2024educational}. The Diagram of a Finite State Machine was also used to test AI captioning systems \cite{buzzi2024isgenerative}. 
        \item Architecture schemas. The accessibility evaluation included the electric car schema and client-server architecture \cite{leotta2024evaluating,pedemonte2025improving}. 
        \item Node-link diagram. These complex schema, often representing structures or relationships, are acknowledged as content types that need further refinement in structured templates, indicating they are still challenging areas of focus \cite{hoppe2021evaluation}. 
    \end{itemize}
    \item General Educational and Informative Images Research also addresses the broader category of images found in educational materials such as illustrations available in textbooks \cite{hoppe2021evaluation,arun2024auditory}. 
\end{itemize}

\subsubsection{Mathematical content}
\label{subsubsec:mathematical_content}
Mathematical representations, characterized by complex notation, symbols, and variables, are addressed through specialized tools focusing on auditory or textual output:
\begin{itemize}
    \item Mathematical expressions and equations. Authors describe this task as challenging due to notation complexity \cite{mondal2019textual,awais2024mathvision}. Research efforts focus on converting mathematical expressions present in images into voice \cite{awais2024mathvision} or generating natural language descriptions that interpret their internal meaning \cite{mondal2019textual,awais2024mathvision}. Sub-categories studied include limits, trigonometric functions, differentiation, and integration \cite{awais2024mathvision}. An easy mathematical equation was also used to evaluate AI tools \cite{buzzi2024isgenerative}. 
    \item Mathematical graphics. Tools and approaches are being explored for creating accessible mathematical diagrams, targeting vector image formats like SVG and XML \cite{leotta2024evaluating}. 
    \item Specific Mathematical Concepts: The Pythagorean Theorem and the Population Pyramid were included in image sets for captioning and accessibility studies \cite{leotta2024evaluating,pedemonte2025improving}. 
\end{itemize}

\subsubsection{Summary of the Findings}
\label{subsubsec:summary_of_the_findings}
Making STEM visual information accessible requires diverse approaches tailored to different content types, ranging from simple charts to complex diagrams and equations. The Table \ref{tab:stem_content} outlines the specific types of visual STEM content targeted in current research and lists the related works.
In summary, while a solid foundation exists in addressing common bar and line charts, the research is progressively moving toward integrating conversational interaction for complex data visualizations, while also tackling the substantial, yet less frequently addressed, challenge of reliably describing complex non-chart STEM images, such as diagrams, schematics, and mathematical equations.

\begin{table}
\caption{Overview of the types of STEM visual content addressed in the reviewed studies.}
\label{tab:stem_content}
\centering

\begin{tabularx}{\textwidth}{X X X}
\hline
\textbf{STEM Image Type}  & \textbf{Specific Examples}  & \textbf{References}  \\
\hline
Data Visualizations (Charts \& Plots)  & Bar Charts, Line Charts, Scatter Plots, Pie Charts  & \cite{shahira2021towards,hoppe2021evaluation,farahani2023automatic,yan2025chart,nylund2025matplotalt}  \\
 & Complex Chart Structures  & \cite{shahira2021towards,gorniak2024vizability,yan2025chart,nylund2025matplotalt,moured2024alt4blind}  \\
\hline
Scientific Diagrams  & General STEM Images  & \cite{shahira2021towards,hoppe2021evaluation,farahani2023automatic,yan2025chart,leotta2024evaluating,pedemonte2025improving,balestrucci2024educational,arun2024auditory,buzzi2024isgenerative}  \\
\hline
Mathematical Content  & Equations and Expressions  & \cite{mondal2019textual,awais2024mathvision,buzzi2024isgenerative}  \\
 & Mathematical Graphics  & \cite{leotta2024evaluating,pedemonte2025improving}  \\
\hline

\end{tabularx}

\end{table}

\subsection{AI Architectures for Description Generation (RQ2)}
\label{subsec:rq2}
The generation of accessible descriptions, often referred to as alternative text (alt-text) or descriptive captions, for visual STEM content relies on AI and machine learning models, particularly those leveraging deep neural networks and multimodal learning. These technologies aim to convert visual data into meaningful textual or auditory content, reducing accessibility gaps for individuals who are blind or have low vision.

At the core of most approaches lies the encoder–decoder architecture, which provides the standard mechanism for converting visual input into natural language. In this framework, the encoder extracts the salient features of an image, typically through a Convolutional Neural Network (CNN). Models such as VGG16, VGG19, ResNet-50, ResNet-152, InceptionV3, and DenseNet are frequently adopted for this purpose, with ResNet-50 often preferred for its ability to capture fine-grained details while mitigating vanishing-gradient issues through its skip-connection design. DenseNet has also proven effective in domains requiring particularly detailed feature extraction, such as images of mathematical expressions. The decoder then processes the encoded feature representation, usually by means of a Recurrent Neural Network (RNN) \cite{elman1990finding}, most notably an LSTM \cite{hochreiter1997long} or GRU, to generate coherent and contextually appropriate textual descriptions. LSTMs are particularly effective at handling long-range dependencies in sequences, while GRUs have also been used as efficient alternatives. These components are commonly trained together within an end-to-end learning paradigm \cite{farahani2023automatic}.

Beyond this foundational structure, several advanced deep learning mechanisms have been introduced to improve description quality and semantic fidelity. Attention mechanisms, for instance, enable the model to focus selectively on the most relevant regions of an image during word generation, addressing limitations related to long-term dependencies and gradient decay observed in traditional RNNs \cite{farahani2023automatic,mondal2019textual,awais2024mathvision,arun2024auditory,karaca2024role}. Sequence-to-sequence (Seq2Seq) modelling, enriched by attention, offers additional flexibility for handling sequences of varying length and has been shown to preserve fine visual details effectively \cite{karaca2024role}. Transformer-based architectures, which rely on self-attention rather than recurrence, are also gaining traction; combinations such as InceptionV3 with a Transformer decoder have demonstrated strong performance in image-to-text generation tasks \cite{farahani2023automatic,karaca2024role,vaswani2017attention}. Other studies explore graph-based approaches, in which images are represented as structured graphs of objects and relations—an effective strategy for modeling semantic dependencies and improving the interpretability of the generated descriptions \cite{farahani2023automatic,karaca2024role}.

Different types of visual STEM content require specific modeling strategies. In the case of data visualizations such as charts and plots, several complementary approaches are used. Some systems rely on LLMs within multi-agent conversational pipelines, where user queries are classified into categories (e.g., analytical, visual, contextual, navigational) and handled by specialized agents such as CSV processors or web-search modules \cite{gorniak2023vizability, gorniak2024vizability}. Other work follows step-by-step pipelines that separate chart-type classification (often using CNNs like VGG16), OCR-based text extraction (e.g., with Tesseract), visual data extraction through rule-based or traditional algorithms, and final description generation using templates or Transformer-based models such as T5 \cite{farahani2023automatic,yan2025chart}. More recent research moves toward end-to-end Vision-Language Models (VLMs)—including MATCHA, UniChart, ChartLlama, ChartAssistant, and TinyChart—which process chart images and natural-language queries directly, often incorporating specialized pre-training tasks such as chart derendering or mathematical reasoning \cite{yan2025chart}.

For mathematical expressions and equations, the literature frequently adopts hybrid models. Systems like MathVision convert equation images into spoken descriptions using Text-to-Speech pipelines \cite{awais2024mathvision}, while others focus first on detecting and classifying equation types with object-detection models such as YOLOv7, distinguishing categories like limits, trigonometry, differentiation, or integration \cite{awais2024mathvision}. Caption-generation models in this domain typically combine CNN encoders (e.g., fine-tuned ResNet-152) with attention-enhanced LSTM decoders to produce natural-language descriptions capable of interpreting the mathematical structure and meaning embedded in the visual notation \cite{mondal2019textual,awais2024mathvision}.

Finally, more general STEM-related educational and scientific images are often handled through custom tools that employ LLM APIs in a black-box fashion to generate alternative descriptions at varying verbosity levels directly during web browsing \cite{pedemonte2025improving}. Studies also evaluate the effectiveness of mature commercial solutions—including Seeing AI, Be My Eyes, and Microsoft Copilot—for their suitability in describing complex STEM content \cite{buzzi2024isgenerative}.

\subsection{The Ecosystem of Datasets and Evaluation (RQ3)}
\label{subsec:rq3}
The quality of generated accessible content, such as image captions or alt text, is evaluated using a combination of automatic metrics, human evaluations, and specialized datasets and benchmarks. These evaluation criteria are often adapted or customized to address the complex and specialized nature of STEM content. However, existing studies indicate a notable divergence between what current computational measures capture and the elements that truly support blind and low-vision users, particularly in terms of delivering relevant visual features in a way that users can clearly interpret. 

\subsubsection{Dataset and Benchmarks}
\label{subsubsec:datasets_and_benchmarks}
A recurring theme across the literature is the critical scarcity of large-scale, high-quality, publicly available datasets specifically designed for STEM image accessibility.
Many studies resort to creating their own small-scale datasets based on the computer vision community that were not originally intended for accessibility applications.
This lack of standardized benchmarks makes it difficult to compare the performance of different models and hinders progress in the field. The creation of such a resource, featuring diverse STEM visuals with rich, multi-level annotations, represents a major need \cite{buzzi2024isgenerative}.
Table \ref{tab:summary_datasets} provides an overview of widely used general-purpose and STEM-specific datasets, detailing the types of images they contain, their approximate size, and the studies that employ them. 
It highlights the predominance of chart-focused synthetic datasets and the relative scarcity of large-scale resources tailored to scientific diagrams, mathematical expressions, or accessibility-oriented annotation.

\begin{table}
\centering
\caption{Summary of the main datasets used for training and evaluating image description systems in STEM contexts.}
\label{tab:summary_datasets}

\begin{tabularx}{\textwidth}{ l X X X}
\hline
Dataset & Image type & \# images  & References \\
\hline
    Flickr8K & General images & 8K & \cite{arun2024auditory,karaca2024role}\\
\hline
MSCOCO & General images & 400k & \cite{ingavelez2022automatic} \\
\hline
DVQA & Bar, Line, Pie charts & 300k & \cite{shahira2021towards,farahani2023automatic}  \\
\hline
FigureQA & Bar, Line, Pie charts & 100k & \cite{shahira2021towards,farahani2023automatic} \\
\hline
FigureSeer & Bar, Line charts & 60k & \cite{shahira2021towards,farahani2023automatic} \\
\hline
LEAF-QA & Bar, Line charts & 250k & \cite{shahira2021towards,farahani2023automatic} \\
\hline
FigCap & Bar, Line charts & 100k & \cite{farahani2023automatic} \\
\hline
SciCap & Bar, Line charts & 400k & \cite{farahani2023automatic} \\
\hline
LineCap & Line charts & 3528 & \cite{farahani2023automatic} \\
\hline
Own dataset & Variable & Variable, in general small  & \cite{gorniak2023vizability,hoppe2021evaluation,leotta2024evaluating,pedemonte2025improving,awais2024mathvision,buzzi2024isgenerative,kim2021information,song2024real,moured2024alt4blind} \\
\hline
Synthetic dataset & Bar, Line charts &  250k & \cite{xie2023prompt} \\
\hline
Math-Exp-Syn & Mathematical expression & 370k & \cite{mondal2019textual} \\
\hline
PlotQA & Bar, Line, Pie charts & 224k & \cite{shahira2021towards} \\
\hline
SythText & Scene-text images generated & 800k & \cite{shahira2021towards} \\
\hline
VisText & Barchart, Area, Line charts & 12k & \cite{nylund2025matplotalt} \\
\hline

\end{tabularx}

\end{table}

\subsubsection{Evaluation Metrics}
\label{subsubsec:evaluation_metrics}
Evaluation metrics fall broadly into two categories: automated metrics (which primarily assess textual similarity) and human-based or advanced metrics (which assess semantic quality, correctness, and usefulness). 

\paragraph{Automated Evaluation Metrics}
To quantify performance, many studies employ automated metrics borrowed from the machine translation and general image captioning fields \cite{farahani2023automatic,yan2025chart,pedemonte2025improving,mondal2019textual,awais2024mathvision,karaca2024role}. As shown in Table \ref{tab:evaluation_metrics}, these include BLEU, ROUGE, METEOR, CIDEr, SPICE, and BERTScore.

\begin{table}
\caption{A summary of the automatic evaluation metrics commonly used in the literature to assess the quality of generated text, highlighting their primary focus, evaluation criteria, and corresponding references.}
\label{tab:evaluation_metrics}
\centering
\begin{tabularx}{\linewidth}{ l X X X }
\hline
Metric  & Focus  & Details and use  & References  \\
\hline
BLEU  & Word Overlap  & Measures the overlap of word sequences (n-grams) between the generated text and a set of human-created reference sentences.  & \cite{farahani2023automatic,yan2025chart,pedemonte2025improving,mondal2019textual,awais2024mathvision,karaca2024role}  \\
\hline
ROUGE  & Content Overlap  & Evaluates generated text by measuring how much of the reference text is covered (overlapping n-grams) by the generated text.  & \cite{farahani2023automatic,yan2025chart,pedemonte2025improving,mondal2019textual,awais2024mathvision,karaca2024role}  \\
\hline
METEOR  & Semantic Similarity  & Improves upon BLEU by incorporating synonym matching and stemming. It showed improved correlation with human judgment.  & \cite{yan2025chart,pedemonte2025improving,karaca2024role}  \\
\hline
CIDEr  & Consensus  & Measures consensus by calculating term frequency-inverse document frequency (TF-IDF) weighted n-gram similarity against multiple reference captions.  & \cite{farahani2023automatic,yan2025chart,pedemonte2025improving,mondal2019textual,karaca2024role}  \\
\hline
SPICE  & Semantic Quality  & Specifically proposed for image captioning, based on the similarity of scene graphs. It analyzes objects, attributes, and relations, correlating better with human judgment for semantic quality.  & \cite{farahani2023automatic}  \\
\hline
BERTScore  & Semantic Evaluation  & BERTScore averages cosine similarity between BERT token embeddings to measure semantic similarity.  & \cite{yan2025chart,nylund2025matplotalt}  \\
\hline

\end{tabularx}

\end{table}

The Bilingual Evaluation Understudy (BLEU) metric evaluates how closely a generated text matches one or more human-written reference sentences by comparing the overlap of word sequences, known as n-grams. A higher score (closer to 100) indicates greater similarity to reference translations. However, it may encourage short sentences and ignores word meaning/synonyms. It is used extensively in image captioning and mathematical equation description tasks. 

The Recall-Oriented Understudy for Gisting Evaluation (ROUGE) metric measures the overlap between the generated and the reference texts based on recall of n-grams, word sequences, and word pairs. Unlike BLEU, which focuses on precision, ROUGE emphasizes how much of the reference content is covered by the generated output. It is particularly useful for evaluating tasks that prioritize completeness of information, such as summarization or detailed image descriptions \cite{yan2025chart,mondal2019textual}. 

The Metric for Evaluation of Translation with Explicit ORdering (METEOR) combines precision, recall, and a harmonic mean (F-score), incorporating stemming, synonymy, and paraphrase matching. This allows to better capture semantic equivalence than BLEU or ROUGE. METEOR accounts for word order by applying a fragmentation penalty, allowing it to better capture fluency and meaning preservation in generated captions \cite{pedemonte2025improving}. 

The Consensus-based Image Description Evaluation (CIDEr) metric is designed for image captioning tasks as it computes consensus between candidate and reference captions by weighting n-grams according to their Inverse Document Frequency (IDF). In this way, this metric gives more importance to informative and less frequent words. CIDEr correlates with human judgement and is a standard metric in visual description benchmarks. However, it may truncate verbose descriptions, compromising comprehensiveness \cite{pedemonte2025improving}. 

The Semantic Propositional Image Caption Evaluation (SPICE) metric focuses on semantic content rather than word overlap. It parses captions into a graph that represents objects, attributes, and relationships, then computes an F-score over the matching semantic tuples \cite{farahani2023automatic}. SPICE is particularly effective in evaluating whether a generated caption is capturing the semantic meaning of the image. However, it may be less sensitive to grammatical fluency \cite{farahani2023automatic}. 

The BERTScore metric leverages contextualized embeddings from pre-trained transformer models to compute similarity between candidate and reference captions at the token level. By aligning words based on their semantic embeddings rather than exact matches, it captures deeper meaning. BERTScore has shown strong correlation with human evaluation \cite{nylund2025matplotalt}. 

These metrics are derived from machine translation and are widely adopted also in image captioning due to their speed and objectivity. However, they have several limitations in assessing semantic quality for complex tasks \cite{farahani2023automatic,yan2025chart,pedemonte2025improving}. 

\paragraph{Human-Centered Evaluation}
Recognizing the limitations of automated metrics, a growing number of studies incorporate human-centred evaluation, as automated metric do not account for usefulness. This is a crucial component for assessing true accessibility. These evaluations can be categorized as follows: 
\begin{itemize}
    \item Custom quality metrics (Correctness, usefulness, completeness). Authors defined a set of custom metrics to assess quality of captions for specialized STEM images \cite{pedemonte2025improving}. These metrics involve dividing the generated description into linguistic predicates (tokens) and manually scoring them: 
    \begin{itemize}
        \item Correctness: measures if the information provided by a token is accurate or inaccurate \cite{pedemonte2025improving}. 
        \item Usefulness: measures if the information is useful for BLV users \cite{pedemonte2025improving}. 
        \item Completeness: measures the proportion of information present in a human defined reference description that is also present in the generated description \cite{pedemonte2025improving}. 
        \item Composite quality metrics (Qua1 and Qua2): these metrics combine the above elements to provide an overall quality score ranging from 0 to 1 \cite{pedemonte2025improving}. Qua1 combines correctness, completeness and incorrectness. $Qua1 = Cor \times Com \times (1 - Inc)$. Where $Qua1 = 1$ is achieved only if the description contains the maximum number of correct tokens. Qua2 incorporates usefulness along with correctness, completeness, and incorrectness: $Qua2 = (Cor AND Use) \times Com \times (1 - Inc)$ \cite{pedemonte2025improving}. 
    \end{itemize}
    \item Human Evaluation (Likert Scales and Preferences): Human raters (often students, developers, or visually impaired participants) are recruited to assess quality subjectively \cite{yan2025chart,leotta2024evaluating,pedemonte2025improving,kim2021information}. 
    \begin{itemize}
        \item 5-Point Likert Scales: Used to assess perceived correctness, usefulness, and quality (ranging from "Very Poor" to "Very Good"). For example, studies assessing LLM performance in generating descriptions for data visualizations, or STEM images, used human raters to score responses based on coherence to ground truth \cite{gorniak2024vizability, hoppe2021evaluation, yan2025chart}. 
        \item Preference Scores: Participants select which generated summary they prefer when presented alongside human-authored or other machine-generated summaries \cite{yan2025chart}. 
        \item Factual Accuracy/Error Categorization: Qualitative analysis is performed by manually labelling and counting error types in generated descriptions, such as value errors, identity errors, trend errors, label errors, missing context, and deceptive errors (hallucinations) \cite{gorniak2024vizability}. 
    \end{itemize}
    \item LLM-based Automatic Evaluation: Large Language Models (LLMs) are increasingly leveraged to perform automatic evaluations often instructed to assess response quality using Likert scales based on coherence to ground truth. For instance, GPT-4 has been used to assess response quality on a 5-point Likert scale, yielding a strong correlation with human assessments (Kendall's $\tau$ score of 0.55) \cite{gorniak2024vizability}. 
\end{itemize}

\paragraph{Reliance on Automated Metrics vs. Human Evaluation}
The primary weakness in current validation methodologies lies in the heavy reliance on automatic metrics derived from machine translation \cite{farahani2023automatic}. 

Metrics like BLEU, ROUGE, and CIDEr are inherently limited because they are based on n-gram matching and only consider word matches, ignoring the meaning of words \cite{yan2025chart,nylund2025matplotalt,pedemonte2025improving}. This results in a very low correlation between these metrics and actual human judgment of quality \cite{farahani2023automatic}. For instance, BLEU captures high overlap in titles and labels but fail entirely to reflect the true semantic meaning or correctness \cite{yan2025chart}. 

These metrics are susceptible to bias and are highly dependent on the quality and format of the reference descriptions in the test set. If the ground truth is flawed, the measured performance can be unreliable. When evaluation metrics that rely on comparing to reference captions, they often reflect the preferences of sighted annotators, not those of blind or low-vision people who actually rely on image descriptions. Hence, the metrics may undervalue descriptions that are more useful or natural for blind users, because they differ in style or content from the sighted-written references \cite{yan2025chart,nylund2025matplotalt}. 

Therefore, human-centred evaluation is indispensable because it evaluates the generated content based on factors critical for application, such as perceived correctness, usefulness, informativeness, and coherence. For complex and high-stakes content, like STEM image descriptions, a human-based evaluation validates if the descriptions adequately serve their target audience, for example, visually impaired students \cite{yan2025chart,leotta2024evaluating,pedemonte2025improving}. 

\subsection{From Static Captions to Multimodal and Interactive Description Systems}
\label{subsec:descrpitive}
A review of image description methods reveals that there are several types of descriptions, ranging from traditional static alt text to multimodal systems.
In Table \ref{tab:descriptions_modalities}, we present a taxonomy of the description methods discussed, highlighting their characteristics, typical use cases, and representative examples from recent studies.

\begin{table}
\caption{Overview of description modalities for accessible STEM images.}
\label{tab:descriptions_modalities}
\centering

\begin{tabularx}{\linewidth}{ X  X  X  X }
\hline
\textbf{Type} & \textbf{Detail and Use} & \textbf{Examples} & \textbf{References} \\
\hline
\textbf{Static (manual/assisted)} & Alt‑text written manually or supported by UI;  & WCAG/W3C guidelines, \textbf{Alt4Blind} (UI + CLIP retrieval), \textbf{MatplotAlt} (Python library) & \cite{nylund2025matplotalt,moured2024alt4blind} \\
\hline
\textbf{Automatic (AI captioning, assistive use)} & Automatically generated descriptions from images/charts; fast but prone to inaccuracies. Used in assistive contexts for STEM and education. & CNN+LSTM, OCR, Transformer (T5, Vision-Language models), Chart2Text, VisText, Seq2Seq with Attention,\textbf{Automated STEM captioning} & \cite{hoppe2021evaluation, yan2025chart, leotta2024evaluating,mondal2019textual,ingavelez2022automatic,arun2024auditory,karaca2024role,song2024real} \\
\hline
\textbf{Dynamic (real‑time, assistive dialogue)} & Adaptive descriptions that update word by word or in real‑time; & VGG19, YOLOv8 (detection+tracking) + T5 (captioning), \textbf{Educational Dialogue System} & \cite{balestrucci2024educational,song2024real} \\
\hline 
\textbf{Multimodal (text + voice + navigation, assistive agents)} & Conversational interaction with charts, sonification, haptics; integrates multiple sensory modalities. Often used in assistive agents for STEM learning. & \textbf{VizAbility} (LLM + Olli Treeview + CSV/Web agents), audio/haptic systems, \textbf{MathVision} (Accessible Intelligent Agent) & \cite{shahira2021towards}, \cite{gorniak2023vizability,gorniak2024vizability,farahani2023automatic,awais2024mathvision,buzzi2024isgenerative,kim2021information} \\
\hline

\end{tabularx}

\end{table}

\subsubsection{Static Captions}
Static descriptions, such as manually created alt text according to WCAG guidelines \cite{yan2025chart}, represent the oldest established practice. Alternative Text (Alt-Text) is the base for web accessibility for images, acting as the primary method for BVI users to acquire information about images through screen readers \cite{shahira2021towards,yan2025chart,pedemonte2025improving,moured2024alt4blind}. Screen readers are unable to interpret non-textual elements such as charts or images, alt-text are normally invisible in visual presentation but are acoustically conveyed by screen readers \cite{farahani2023automatic}. However, traditional alt-text is often a simple caption characterized as a static text dump because it tends to be passive, unstructured, generic, and short, frequently lacking sufficient relevant information or violating accessibility guidelines \cite{leotta2024evaluating,moured2024alt4blind}. This can result in oversimplification or complications for the user \cite{moured2024alt4blind}.
For complex visual content, such as figures in educational or scientific materials (STEM images), a simple alt-text is insufficient \cite{leotta2024evaluating,pedemonte2025improving}. Comprehensive descriptions are necessary, requiring:
\begin{enumerate}
    \item Completeness and Correctness: Descriptions must contain complete and accurate information \cite{hoppe2021evaluation}.
    \item Context: Used symbols should be described with their semantic meaning, not just their shape \cite{hoppe2021evaluation}.
    \item Structure and Granularity: Descriptions should allow the user to navigate varying levels of detail \cite{hoppe2021evaluation,leotta2024evaluating}.
    \item Clarity: Descriptions must be written in clear, understandable language, avoiding unfamiliar jargon or overly technical terms \cite{leotta2024evaluating,pedemonte2025improving}.
\end{enumerate} 
Templates and structured formats are used to address the challenge of generating detailed descriptions, especially for plots and diagrams \cite{hoppe2021evaluation,nylund2025matplotalt}. A library named Olli allows to transform charts into a keyboard-navigable text representation (tree view) \cite{gorniak2024vizability}.
Static captions while essential, they are often insufficient to convey the complexity of STEM diagrams and equations. 

\subsubsection{Automatic Captioning}
Moving beyond static text, several systems incorporate interactive, multimodal, and conversational elements to provide a rich, user-driven experience for exploring visual data.
These assisted frameworks such as Alt4Blind \cite{moured2024alt4blind} and MatplotAlt \cite{nylund2025matplotalt} extend this paradigm by simplifying or automating graph descriptions. More recent developments concern automatic captioning, where large-scale text summaries are produced by models such as Convolutional Neural Network (CNN) + Long Short‑Term Memory (LSTM) \cite{arun2024auditory}, Optical Character Recognition (OCR) pipelines, and Transformer-based architectures (T5, GPT variants) \cite{mondal2019textual,karaca2024role,song2024real}, although accuracy in technical domains is still an issue. The most promising frontier is represented by multimodal systems, which integrate textual, audio, tactile, and conversational interfaces to allow blind and visually impaired users to explore graphs and scientific equations through multiple sensory channels. This convergence is demonstrated by frameworks such as VizAbility \cite{gorniak2023vizability,gorniak2024vizability}, haptic and audio systems \cite{shahira2021towards,farahani2023automatic,kim2021information}, and intelligent agents such as MathVision \cite{awais2024mathvision}, which often overlap with educational and assistive functions \cite{balestrucci2024educational,ingavelez2022automatic,song2024real}. Dynamic approaches introduce adaptability: real-time captioning systems (YOLOv8 + T5) \cite{song2024real} enable contextualized and editable explanations, particularly useful in educational contexts. Finally, analytical and evaluative studies such as Prompt Log Analysis \cite{xie2023prompt} and dataset benchmarking provide crucial insights into the limitations and quality of the generated descriptions. In this direction, multimodal and assistive systems hold the greatest promise for future accessibility in STEM fields, while static descriptions now seem obsolete. 

\subsubsection{Effective Description}
An effective description for a BVI user must be structured to facilitate the construction of a mental model. The literature provides a robust framework for this, most notably the four-level model of semantic content proposed by Lundgard and Satyanarayan in 2021 \cite{lundgard2021accessible}:
\begin{itemize}
    \item Level 1 (L1): Elemental and Encoded Information. This covers the chart type, title, axes labels, and legends.
    \item Level 2 (L2): Statistical and Relational Information. This describes explicit data facts, such as minimum/maximum values or relationships between data points.
    \item Level 3 (L3): Perceptual and Cognitive Information. This captures the primary insights a sighted user would perceive, such as overall trends, patterns, and outliers.
    \item Level 4 (L4): Contextual and Domain-Specific Information. This provides external context or a broader narrative related to the chart's subject matter.
\end{itemize}
User studies consistently demonstrate that BVI users find descriptions combining these levels to be most effective, preferring a structure that starts with L1 information before moving to L2 or L3 details. L4 content is often considered the least useful, as users prefer factual data over background information \cite{lundgard2021accessible}.

\begin{table}
\centering
\caption{Summary of empirical models for structuring effective alt text for STEM visualizations.}
\label{tab:empiriacal_models}

\begin{tabularx}{\linewidth}{ X  X  X }
\hline
Paper & Alt Text Content Assessment & Alt Text Content Conclusion \\
\hline
\cite{lundgard2021accessible} & Four-level model (L1-L4) & Visually impaired people prefer combined levels of information (e.g., L1+L2/L3) and favor L3 over L2. L4 is the least useful. \\
\cite{zong2022rich} & A tree structure of information. & A hierarchical structure is effective with minimal cognitive load. A data table is necessary for accessing individual data points. \\
\cite{kim2021accessible} & 'Existence', 'Overview', and 'Details' categories. & Descriptions should begin by indicating a chart is present, followed by an overview of visual encoding and key insights.  \\
\hline

\end{tabularx}

\end{table}

Other proposed models show a similar emphasis on progressively structured information, reinforcing the idea that effective descriptions must balance completeness with cognitive accessibility, as shown in Table \ref{tab:empiriacal_models}. Despite differences in terminology, these frameworks converge on three key principles: (1) users benefit from an initial global orientation before receiving specific data facts; (2) summaries of perceptual insights, such as trends or notable features, are essential for building a meaningful mental model; and (3) excessive contextual or narrative information tends to hinder rather than support comprehension. Taken together, these findings highlight the need for layered, insight-driven description strategies that foreground visually salient information while providing optional access to fine-grained details when required by the user.

\subsection{Open Problems and Emerging Research Opportunities (RQ5)}
\label{subsec:open_problems}
The current trajectory toward making STEM visual information accessible through AI faces several significant technical and practical limitations. Researchers propose charting a path forward by focusing on robust multimodal integration, high-quality specialized data creation, and enhanced user agency and customization. 

\subsubsection{Current Limitations and Constraints}
\begin{figure}
    \centering
    \includegraphics[width=1\linewidth]{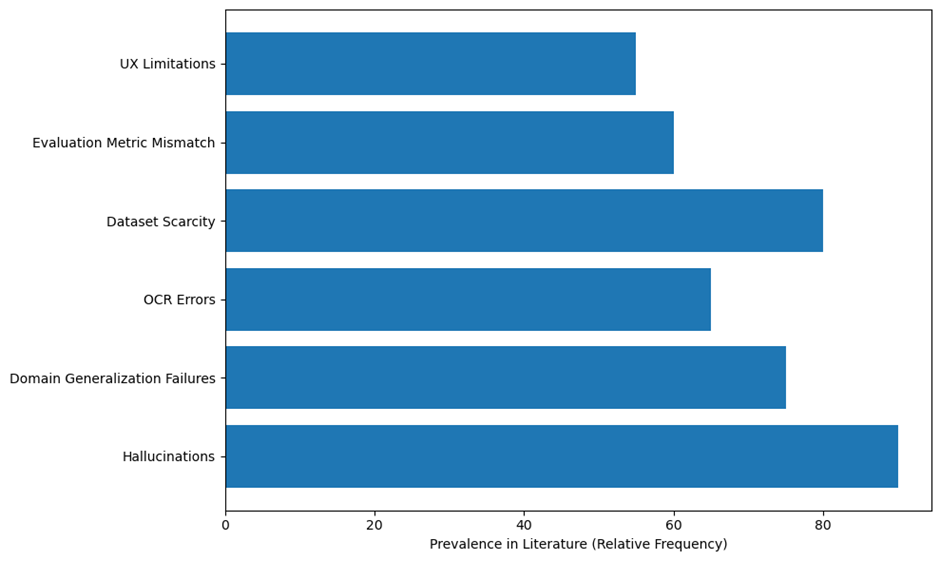}
    \caption{Relative prevalence of common limitations reported in the literature on AI-generated descriptions of STEM visuals.}
    \label{fig:limitations_categories}
\end{figure}

Figure \ref{fig:limitations_categories} summarizes the principal categories of limitations identified across the reviewed studies, including hallucinations, domain generalization failures, OCR-related errors, data scarcity, metric misalignment, and computational constraints.
Current approaches face constraints related to model accuracy, the lack of specialized data, and challenges in deployment and adoption. Among these, AI-generated descriptions, particularly those from Vision-Language Models and LLMs, remain susceptible to factual errors and hallucinations \cite{nylund2025matplotalt,pedemonte2025improving,buzzi2024isgenerative}. These inaccuracies drastically reduce the perceived correctness, usefulness, and quality of descriptions, especially for complex STEM images \cite{leotta2024evaluating}. In competitive evaluations, AI-generated descriptions for STEM content received lower quality ratings compared to human-written descriptions \cite{leotta2024evaluating}. For instance, one proprietary VLM model, tested out-of-the-box, had captions where 81 out of 100 contained some kind of factual inaccuracy \cite{nylund2025matplotalt}. Open-source models like ViT-GPT and BLIP struggle with complex STEM diagrams (e.g., anatomy), exhibiting severe hallucinations or lacking the necessary level of anatomical detail required for educational context \cite{pedemonte2025improving}. Similarly, LLMs tested on describing FSA proved inadequate in accuracy and consistency of responses, sometimes making errors in syntax or providing incomplete explanations \cite{balestrucci2024educational}.
Similarly, AI models suffer for poor domain generalization capabilities. The general image captioning models, designed primarily for natural scenes, cannot be reliably applied to domain-specific images like scientific charts and graphs \cite{farahani2023automatic}. Automating chart understanding is complicated by chart type variability, as designing a single unified model that handles different chart types (bar, line, scatter, node-link diagrams) accurately is difficult \cite{shahira2021towards, farahani2023automatic}. Existing template-based or modular systems (e.g., AutoCaption, ChartVi) often rely on pre-defined templates, which leads to stereotyped sentences lacking generality \cite{yan2025chart}. Models that perform well on simple chart types (like single-lined figures) often struggle with complex trends or multi-lined charts, sometimes repeating the same description for multiple data series \cite{farahani2023automatic}.
Generating accurate responses in chart QA heavily relies on initial image processing steps like OCR, where extraction errors (e.g., failing to recover data values) can lead to sub-optimal and inaccurate results \cite{gorniak2024vizability,yan2025chart}. Accurate visual processing is hindered by problems such as low image quality, complex chart designs, occlusion, and variations in font styles and sizes \cite{shahira2021towards,farahani2023automatic}. Current chart QA systems face difficulties with complex reasoning questions that require combining visual elements with implied relationships, as models must go beyond simply retrieving textual data to identify relationships or complex calculations (e.g., area-under-the-curve) \cite{shahira2021towards,farahani2023automatic}. The large scale of modern language and vision models (LLMs/VLMs) necessitates significant computational costs, making it challenging to guarantee real-time performance, which is crucial for dynamic interactions like interpreting live lectures or video conferences \cite{song2024real}.
A major challenge is the limited availability of large-scale, high-quality, publicly available datasets specifically tailored for scientific chart image captioning and accessibility tasks \cite{farahani2023automatic,yan2025chart}. Most available datasets, presented in Section \ref{subsubsec:datasets_and_benchmarks}, are synthetically created or lack verified ground truth alternative texts that align with the specific needs of visually impaired users.
Standard automated evaluation metrics (e.g., BLEU, ROUGE, CIDEr), described in Section \ref{subsubsec:evaluation_metrics} are not reliable for assessing accessibility-focused descriptions in complex domains like STEM. These metrics prioritize exact lexical match, often penalize the necessary verbosity, and fail to capture semantic richness or the importance of contextual accuracy \cite{farahani2023automatic, pedemonte2025improving}. Current accessibility guidelines often focus on simple charts, and there is a lack of empirical evidence and standardization for complex, grouped, or scientific charts, where simple alt text cannot convey sufficient insights concisely \cite{yan2025chart}.
While combining interaction modalities (like keyboard navigation and conversational AI, as in VizAbility) enhances accessibility, complex interfaces or new workflows may be difficult for screen reader users to adopt \cite{gorniak2023vizability,yan2025chart}. Traditional "one-size-fits-all" alt-text approaches do not meet diverse user preferences. The need for user-driven customization regarding description length, verbosity, and content ordering is a pivotal yet frequently unaddressed design consideration \cite{gorniak2024vizability,nylund2025matplotalt,pedemonte2025improving}. The reliance on LLMs introduces risks related to bias, fairness, and the potential for misleading users if generated errors are overlooked, especially in high-stakes fields like medicine or education \cite{nylund2025matplotalt,pedemonte2025improving}.

\subsubsection{Future Research Directions}
Future research aims to overcome these limitations by improving models, creating better resources, and focusing heavily on user-centric design and validation:
\begin{itemize}
    \item Enhancing AI Architectures and Reasoning. Research should focus on integrating emerging vision capabilities with the symbolic, data-driven reasoning power of specialized LLM pipelines to enhance robustness and accuracy \cite{gorniak2024vizability,farahani2023automatic,pedemonte2025improving}. Future systems must be developed to handle multi-hop reasoning and analyze complex natural language questions that require interpreting relationships and patterns not explicitly encoded in the raw data \cite{shahira2021towards,gorniak2024vizability, farahani2023automatic}. For educational contexts, developing more explainable and manageable dialogue systems is preferable to adopting opaque black-box LLMs, which lack precision control over content and interaction dynamics necessary for teaching \cite{balestrucci2024educational}. Strategies like prompting VLMs with heuristic alt text and data tables should be further explored and optimized, as this technique has been shown to increase the factuality of generations and reduce deceptive errors \cite{nylund2025matplotalt}. Development should continue on efficient architectures, possibly utilizing detection-captioning-tracking pipelines, to guarantee real-time performance in dynamic environments like live videos or online classes, especially when computational resources are limited \cite{song2024real}.
    \item Developing New Data and Evaluation Resources. There is an urgent need to construct larger, more realistic, and inclusive benchmark datasets that incorporate complex and grouped chart types \cite{gorniak2024vizability,yan2025chart}. These datasets must be verified using the viewpoints and real-world questions posed by blind and low-vision users to ensure their relevance \cite{gorniak2024vizability}. To address the scalability issue of manual quality assessment, future work should investigate utilizing LLMs themselves to assist in evaluating descriptions, assessing metrics like fluency, informativeness, and coherence, either as initial filters or within a human-in-the-loop validation process \cite{gorniak2024vizability,pedemonte2025improving}. Detailed analysis of existing descriptions should be conducted to create a "ground truth" description for STEM images that fully complies with specifications for correctness and usefulness, facilitating the creation of objective benchmarks \cite{leotta2024evaluating}. Comparisons between different AI engines focusing specifically on STEM descriptions are also necessary \cite{leotta2024evaluating}.
    \item Focusing on User-Centric Design and Deployment. Future tools should prioritize functionality allowing users to customize verbosity levels, description ordering, and content to accommodate diverse individual needs and preferences \cite{gorniak2024vizability,nylund2025matplotalt,pedemonte2025improving}. This involves integrating controls for developers and users (e.g., changing node descriptions in interactive charts) \cite{gorniak2024vizability}. Researchers must conduct rigorous user studies involving larger and more diverse groups of BLV participants to thoroughly assess the practical utility and usability of systems \cite{gorniak2023vizability,gorniak2024vizability,balestrucci2024educational}. This includes investigating the detail-accuracy trade-off and the level of trust BVI users place in potentially error-prone AI-generated descriptions \cite{nylund2025matplotalt}. Efforts should explore integrating tools into existing visualization tools, web browsers, and educational platforms (like Content Management Systems) to allow for real-time alt-text generation during content creation, promoting accessibility by design \cite{gorniak2024vizability,yan2025chart,pedemonte2025improving}. Future research should explore expanding accessibility solutions to accommodate multiple-coordinated and dynamic data visualizations (like dashboards), investigating how conversational agents can resolve ambiguities in such complex environments \cite{gorniak2024vizability}.
\end{itemize}

\section{Discussion}
\label{sec:discussion}
The accessibility of STEM content—particularly graphical and visual materials such as charts, diagrams, and scientific illustrations—has become an increasingly relevant field to assure equal educational opportunity for visually-impaired students. Although the initial search retrieved a considerable number of articles, the final review included only twenty studies. This indicates that the field is still in an emerging phase. The small corpus and its temporal distribution, spanning the period from 2019 to 2025, suggest a scientific community that is growing but still immature in addressing accessibility support for visual content in STEM domains. Recent initiatives, although numerically more numerous, remain largely focused on the technical demonstration of new artificial intelligence models rather than on assessing their effectiveness and usability for blind or visually impaired users.
From a technological perspective, a certain discontinuity can be observed: many studies adopt traditional machine learning architectures (such as CNNs) that, at the time of publication, had already been surpassed by more advanced multimodal approaches. This highlights how rapidly the field of AI is evolving.
Another key finding concerns the lack of adequate datasets for STEM-related images to generate effective and meaningful alternative descriptions \cite{cardia2025descriptive}. The absence of large-scale, high-quality resources designed for accessibility represents not only a technical limitation but also a lack of attention to and research on inclusion. Without such datasets, AI models are trained exclusively on visual and textual data that overlook the needs of people who access images through alternative modalities. Many of the reviewed studies rely on generic or small datasets, often originally created for general image captioning. Most articles focus on quantitative visual representations—such as bar charts—while studies dealing with structural or conceptual elements typical of scientific disciplines (e.g., graphs, flow diagrams, or relational structures) are rare. At present, there are no datasets specifically designed to generate truly accessible descriptions, whether for general understanding of STEM images or for educational purposes. This absence limits the possibility of assessing the performance of automatic generation systems in a comparable and reproducible manner, preventing the definition of shared benchmarks.
Another significant aspect that emerged from the review is the near-total absence of explicit references to the Web Content Accessibility Guidelines (WCAG). None of the included studies make direct reference to the W3C success criteria, not even the most basic requirement 1.1.1 on “non-text content.” At most, there are generic mentions of the need to provide alternative text. Addressing this gap would require dedicated systematic research specifically focused on web accessibility practices and standards.

\subsection{Summary of Findings}
In the following, we summarize the relevant results for each RQ:
\begin{itemize}
    \item \textbf{RQ1} — Types of visual STEM content targeted
The review shows that current research overwhelmingly concentrates on a narrow subset of STEM visual content, with data visualizations, in particular bar charts, line charts, scatter plots, and pie charts, receiving the greatest attention. These chart types are predominant in public datasets due to their relatively simple structure and due to the possibility of synthetic generation. On the other hand, scientific diagrams and flowcharts appear less frequently in the literature. This imbalance highlights the presence of a gap between research focus and the diversity of visual artifacts used in real STEM communication, indicating that accessibility solutions remain limited and do not yet address the user needs.
    \item \textbf{RQ2} — AI/ML approaches for description generation
The surveyed studies present an evolution in the methodology. The early stage approaches relied on multiple stages approaches, rule-based pipelines and CNN-based feature extraction, while more recent systems adopt transformer-based architectures, multimodal Vision-Language Models, and Large Language Models able to generate fluent, context-aware descriptions. Nevertheless, the challenge of ensuring good performance persists across all techniques. Models often produce descriptions that are syntactically coherent but semantically incorrect, misstate relationships between data points, or hallucinate elements that do not exist in the visualization. This indicates that although generative models have become more expressive, they still lack robust mechanisms for grounding descriptions in precise visual data, which is essential for STEM content.
\item \textbf{RQ3} — Evaluation methods, datasets, and metrics
Evaluation practices across the included studies reveal substantial fragmentation and misalignment with accessibility goals. Most works rely on synthetic or general-purpose figure-caption datasets that were not designed to support accessibility research, and only a very small number of datasets include descriptions authored or validated by blind or low vision users. As a result, models are typically assessed using automated text-overlap metrics such as BLEU, ROUGE, METEOR, and CIDEr, which measure surface-level similarity but fail to capture the correctness, completeness, or usability of STEM descriptions for accessibility purposes. Human evaluations remain limited and inconsistent, and studies involving actual BLV users are rare. These issues substantially limit the validity and ecological relevance of current evaluation methodologies.
\item \textbf{RQ4} — Description modalities, interaction models, and accessibility tools
In this review we identified two main paradigms in the design of output modalities: static description generation and interactive systems. Most studies still produce static alternative text or chart summaries, often reflecting minimal compliance with accessibility norms. However, a growing subset of recent work adopts more sophisticated interaction models, including conversational interfaces, keyboard-based navigation, and tiered or multi-level descriptions that allow the user to request additional detail on demand. These interactive approaches mark a shift from passive consumption toward active exploration, enabling BLV users to interrogate visual data more deeply. Nonetheless, such systems are still in early development stages and lack standardization, making it difficult to evaluate their effectiveness or integrate them with mainstream accessibility workflows.
    \item \textbf{RQ5} — Limitations of current approaches and future research directions
Across the literature, persistent limitations affect both the technical and practical dimensions of accessible description generation. On the technical side, models frequently struggle with the accurate extraction and interpretation of numerical values, trends, and relationships, particularly when dealing with complex or multi-series visualizations. On the practical side, progress is hampered by the scarcity of accessibility-first datasets, limited user-centered evaluation, and inconsistent adherence to accessibility standards such as WCAG. These gaps point to several actionable directions for future research, including the development of factually grounded multimodal models, the creation of large-scale datasets co-designed with BLV communities, improved support for complex real-world scientific figures, and the expansion of interactive and multimodal accessibility tools that empower users to navigate and understand STEM visuals independently.
\end{itemize}
While accessibility remains the primary motivation for generating descriptions of STEM visuals, our analysis shows that their impact extends far beyond assistive technologies. Alternative text increasingly functions as machine-readable metadata for search engines, scientific repositories, and multimodal LLMs, enabling more accurate retrieval, classification, and reasoning over visual scientific knowledge. As future vision-language and image-generation models are trained on large-scale STEM corpora, the availability of precise, accessibility-oriented descriptions will become critical for ensuring factual grounding and trustworthy generation. Therefore, improving the quality and standardization of STEM alt-text is not only essential for equitable access but also for supporting the next generation of HCI tools and AI systems that operate over, transform, or generate visual scientific information.

\section{Conclusion}
\label{sec:conclusion}
This systematic survey has explored the AI-powered techniques for making STEM visuals accessible to people who are blind or have low vision. The field is characterized by rapid technological advancement, driven by the promise of leveraging AI to dismantle a significant barrier to educational and professional equity. The analysis reveals a clear trajectory from simple, static descriptions of common chart types towards more sophisticated, interactive systems capable of handling a wider array of complex scientific content.
Despite this progress, the findings underscore a set of profound and interconnected challenges that define the field's current frontier. The gap between superficial pattern recognition and deep semantic reasoning remains wide. The crisis of trust, fueled by the potential for factual inaccuracies and model hallucinations, poses a critical threat to the real-world adoption of these technologies. Finally, the prevailing evaluation paradigms, with their over-reliance on inadequate automated metrics and demographically opaque user studies, hinder robust scientific progress and risk creating solutions that are not genuinely useful or equitable.
The path forward lies not in the pursuit of a single, fully autonomous "magic bullet," but in a more nuanced, human-centered approach. By focusing on building trustworthy, verifiable, and personalized systems, and by embracing a paradigm of Human-AI Collaboration, the research community can create tools that empower, rather than replace, human expertise. By addressing the methodological and ethical shortcomings in evaluation, we can ensure these tools serve the diverse needs of the entire BLV community. Ultimately, the goal of this research extends beyond mere information access; it is about fostering epistemological equity, ensuring that all individuals have the tools to independently engage with, critique, and contribute to the scientific endeavor. The work ahead is substantial, but the potential to create a more inclusive and accessible future for STEM is a powerful and worthy motivation.
Future research opportunities focus on addressing these limitations by creating more robust, integrated, and user-centric AI systems for STEM accessibility. Future work aims to refine models to be more accurate, contextually aware, and reliable, particularly in handling domain-specific visual reasoning. Future research must prioritize strategies for addressing hallucinations and improving the factual accuracy and consistency of AI-generated text, particularly for complex visuals. Incorporating emerging Vision-LLMs (Vision-Large Language Models) is a promising direction, aiming to integrate their visual description strengths with structured, symbolic processing to achieve greater performance and interpret low-level analytic tasks to higher-level cognitive operations. Exploring more sophisticated training techniques, such as reinforcement learning from human feedback (RLHF), could offer richer supervision and guidance, helping to align model output more closely with user needs and accessibility standards
New datasets and evaluation methods are necessary to support the advanced models under development. There is a critical need to construct larger, more rigorous, and inclusive benchmark datasets that incorporate the specific questions and viewpoints of blind and low-vision individuals. New datasets should include multiple types of alternative descriptions to prepare models for real-world scenarios. Researchers should investigate deploying LLMs themselves as tools to assist in evaluating the quality of descriptions (e.g., acting as first-pass filters or within a human-in-the-loop framework) to enable scalable benchmarking across domains.
Moving beyond static descriptions, research will focus on creating truly dynamic and user-driven exploration experiences, by integrating conversational agents into multi-coordinated charts or dashboards, requiring solutions for resolving ambiguities in user queries about relevant charts. Future systems should be designed to handle multi-category questions requiring advanced reasoning (multi-hop reasoning) and situational questions that depend on the user’s current context or location within a navigable chart structure. Research can focus on developing reliable metrics and strategies to actively encourage exploratory generation, allowing users to probe different descriptive possibilities or navigate information spaces efficiently. Integration of dialogue systems or conversational interaction is key for complex structures like FSA and mathematical content, as this approach has been shown to be more effective than traditional tabular descriptions.

\section*{Acknowledgements}

This work was supported by the STEMMA PRIN project thanks to the funding by the European Union - Next Generation EU, Mission 4 Component 2 CUP B53D23019500006.

\section*{Disclosure statement}

The authors declare no conflicts of interest.

\section*{Notes on contributors}
Marco Cardia: Conceptualization; Methodology; Data curation; Formal analysis; Investigation; Visualization; Writing – original draft; Writing – review and editing; Project administration.

Letizia Angileri: Conceptualization; Data curation; Investigation; Writing – review and editing.

Marina Buzzi: Conceptualization; Investigation; Supervision; Writing – review and editing.

Giulio Galesi: Data curation; Formal analysis; Investigation; Validation; Visualization; Writing – review and editing.

Barbara Leporini: Conceptualization; Funding acquisition; Supervision; Validation; Writing – review and editing; Project administration.

All authors contributed substantially to the design of the work and the analysis of the data; they all participated in revising it critically for important intellectual content; they all approved the final version to be published; and they all agree to be accountable for all aspects of the work.

\bibliographystyle{IEEEtran}
\bibliography{interactapasample}

@article{mukhiddinov2021systematic,
  title={A systematic literature review on the automatic creation of tactile graphics for the blind and visually impaired},
  author={Mukhiddinov, Mukhriddin and Kim, Soon-Young},
  journal={Processes},
  volume={9},
  number={10},
  pages={1726},
  year={2021},
  publisher={MDPI}
}

@misc{brinn2022framework,
      title={A framework for improving the accessibility of research papers on arXiv.org}, 
      author={Shamsi Brinn and Christopher Cameron and David Fielding and Charles Frankston and Alison Fromme and Peter Huang and Mark Nazzaro and Stephanie Orphan and Steinn Sigurdsson and Ryan Tay and Miranda Yang and Qianyu Zhou},
      year={2024},
      eprint={2212.07286},
      archivePrefix={arXiv},
      primaryClass={cs.DL},
      url={https://arxiv.org/abs/2212.07286}, 
}

@article{poggianti2025immersive,
  title={Immersive technologies for inclusive digital education: a systematic survey},
  author={Poggianti, Camilla and Chessa, Stefano and Pelagatti, Susanna and Kocian, Alexander},
  journal={Human Behavior and Emerging Technologies},
  volume={2025},
  number={1},
  pages={8888303},
  year={2025},
  publisher={Wiley Online Library}
}

@inproceedings{cardia2025descriptive,
  author = {Cardia, Marco and Buzzi, Marina and Galesi, Giulio and Leporini, Barbara},
    title = {{A Descriptive Review of Image Datasets for Accessible Alternative Descriptions in STEM Domains}},
    year = {2025},
    isbn = {9798400721021},
    publisher = {Association for Computing Machinery},
    address = {New York, NY, USA},
    doi = {10.1145/3750069.3750122},
  pages={1--7},
  booktitle = {Proceedings of the 16th Biannual Conference of the Italian SIGCHI Chapter},
articleno = {9},
numpages = {7},
location = {Salerno, Italy
},
series = {CHItaly '25},
}

@article{shahira2021towards,
  title={Towards assisting the visually impaired: a review on techniques for decoding the visual data from chart images},
  author={Shahira, KC and Lijiya, A},
  journal={IEEE Access},
  volume={9},
  pages={52926--52943},
  year={2021},
  publisher={IEEE}
}

@inproceedings{singh2024figura11y,
  title={{Figura11y: Ai assistance for writing scientific alt text}},
  author={Singh, Nikhil and Wang, Lucy Lu and Bragg, Jonathan},
  booktitle={Proceedings of the 29th International Conference on Intelligent User Interfaces},
  pages={886--906},
  year={2024}
}

@article{lundgard2021accessible,
  title={Accessible visualization via natural language descriptions: A four-level model of semantic content},
  author={Lundgard, Alan and Satyanarayan, Arvind},
  journal={IEEE transactions on visualization and computer graphics},
  volume={28},
  number={1},
  pages={1073--1083},
  year={2021},
  publisher={IEEE}
}

@article{yan2025chart,
  title={{Chart Accessibility: A Review of Current Alt Text Generation}},
  author={Yan, Chuqiao and Hutter, Hans-Peter and Schmitt-Koopmann, Felix M and Darvishy, Alireza},
  journal={IEEE Access},
  year={2025},
  publisher={IEEE}
}

@inproceedings{zong2022rich,
  title={Rich screen reader experiences for accessible data visualization},
  author={Zong, Jonathan and Lee, Crystal and Lundgard, Alan and Jang, JiWoong and Hajas, Daniel and Satyanarayan, Arvind},
  booktitle={Computer Graphics Forum},
  volume={41},
  number={3},
  pages={15--27},
  year={2022},
  organization={Wiley Online Library}
}

@inproceedings{gorniak2023vizability,
  title={{Vizability: Multimodal accessible data visualization with keyboard navigation and conversational interaction}},
  author={Gorniak, Joshua and Ottiger, Jacob and Wei, Donglai and Kim, Nam Wook},
  booktitle={Adjunct Proceedings of the 36th Annual ACM Symposium on User Interface Software and Technology},
  pages={1--3},
  year={2023}
}

@inproceedings{gorniak2024vizability,
  title={{Vizability: Enhancing chart accessibility with llm-based conversational interaction}},
  author={Gorniak, Joshua and Kim, Yoon and Wei, Donglai and Kim, Nam Wook},
  booktitle={Proceedings of the 37th Annual ACM Symposium on User Interface Software and Technology},
  pages={1--19},
  year={2024}
}

@inproceedings{hoppe2021evaluation,
  title={Evaluation of automated image descriptions for visually impaired students},
  author={Hoppe, Anett and Morris, David and Ewerth, Ralph},
  booktitle={International Conference on Artificial Intelligence in Education},
  pages={196--201},
  year={2021},
  organization={Springer}
}

@article{pedemonte2025improving,
  author={Pedemonte, Giacomo and Leotta, Maurizio and Ribaudo, Marina},
  journal={IEEE Access}, 
  title={{Improving Web Accessibility With an LLM-Based Tool: A Preliminary Evaluation for STEM Images}}, 
  year={2025},
  volume={13},
  number={},
  pages={107566-107582},
  doi={10.1109/ACCESS.2025.3577519}
  }

@inproceedings{mondal2019textual,
  title={Textual description for mathematical equations},
  author={Mondal, Ajoy and Jawahar, CV},
  booktitle={2019 International Conference on Document Analysis and Recognition (ICDAR)},
  pages={1300--1307},
  year={2019},
  organization={IEEE}
}

@article{awais2024mathvision,
  title={Mathvision: An accessible intelligent agent for visually impaired people to understand mathematical equations},
  author={Awais, Muhammad and Ahmed, Tauqir and Aslam, Muhammad and Rehman, Amjad and Alamri, Faten S and Bahaj, Saeed Ali and Saba, Tanzila},
  journal={IEEE Access},
  year={2024},
  publisher={IEEE}
}

@article{balestrucci2024educational,
  title={An educational dialogue system for visually impaired people},
  author={Balestrucci, Pier Felice and Di Nuovo, Elisa and Sanguinetti, Manuela and Anselma, Luca and Bernareggi, Cristian and Mazzei, Alessandro},
  journal={IEEE Access},
  year={2024},
  publisher={IEEE}
}

@article{farahani2023automatic,
  title={Automatic chart understanding: a review},
  author={Farahani, Ali Mazraeh and Adibi, Peyman and Ehsani, Mohammad Saeed and Hutter, Hans-Peter and Darvishy, Alireza},
  journal={IEEE Access},
  volume={11},
  pages={76202--76221},
  year={2023},
  publisher={IEEE}
}

@article{ingavelez2022automatic,
  title={Automatic adaptation of open educational resources: an approach from a multilevel methodology based on students’ preferences, educational special needs, artificial intelligence and accessibility metadata},
  author={Ingav{\'e}lez-Guerra, Paola and Robles-Bykbaev, Vladimir E and Perez-Mu{\~n}oz, Angel and Hilera-Gonz{\'a}lez, Jos{\'e} and Oton-Tortosa, Salvador},
  journal={IEEE Access},
  volume={10},
  pages={9703--9716},
  year={2022},
  publisher={IEEE}
}

@inproceedings{arun2024auditory,
  title={Auditory aid for understanding images for Visually Impaired Students using CNN and LSTM},
  author={Arun, Aashna and Sahay, Apurvanand},
  booktitle={2024 15th International Conference on Computing Communication and Networking Technologies (ICCCNT)},
  pages={1--7},
  year={2024},
  organization={IEEE}
}

@article{karaca2024role,
  title={The Role of Attention Mechanism in Generating Image Captions: An Innovative Approach with Neural Network-Based Seq2seq Model},
  author={Karaca, Zeynep and Da{\c{s}}, Bihter},
  journal={Sakarya University Journal of Computer and Information Sciences},
  volume={7},
  number={1},
  pages={92--102},
  year={2024},
  publisher={Sakarya University}
}

@article{elman1990finding,
  title={Finding structure in time},
  author={Elman, Jeffrey L},
  journal={Cognitive science},
  volume={14},
  number={2},
  pages={179--211},
  year={1990},
  publisher={Wiley Online Library}
}

@article{hochreiter1997long,
  title={Long short-term memory},
  author={Hochreiter, Sepp and Schmidhuber, J{\"u}rgen},
  journal={Neural computation},
  volume={9},
  number={8},
  pages={1735--1780},
  year={1997},
  publisher={MIT press}
}

@article{vaswani2017attention,
  title={Attention is all you need},
  author={Vaswani, Ashish and Shazeer, Noam and Parmar, Niki and Uszkoreit, Jakob and Jones, Llion and Gomez, Aidan N and Kaiser, {\L}ukasz and Polosukhin, Illia},
  journal={Advances in neural information processing systems},
  volume={30},
  year={2017}
}

@inproceedings{nylund2025matplotalt,
  title={MatplotAlt: A Python Library for Adding Alt Text to Matplotlib Figures in Computational Notebooks},
  author={Nylund, Kai and Mankoff, Jennifer and Potluri, Venkatesh},
  booktitle={Computer Graphics Forum},
  pages={e70119},
  year={2025},
  organization={Wiley Online Library}
}

@inproceedings{buzzi2024isgenerative,
author={Marina Buzzi and Giulio Galesi and Barbara Leporini and Annalisa Nicotera},
title={Is Generative AI Mature for Alternative Image Descriptions of STEM Content?},
booktitle={Proceedings of the 20th International Conference on Web Information Systems and Technologies - WEBIST},
year={2024},
pages={274-281},
publisher={SciTePress},
organization={INSTICC},
doi={10.5220/0012996800003825},
isbn={978-989-758-718-4},
issn={2184-3252},
}

@inproceedings{kim2021information,
  title={Information graphic summarization using a collection of multimodal deep neural networks},
  author={Kim, Edward and Onweller, Connor and McCoy, Kathleen F},
  booktitle={2020 25th International Conference on Pattern Recognition (ICPR)},
  pages={10188--10195},
  year={2021},
  organization={IEEE}
}

@inproceedings{kim2021accessible,
  title={Accessible visualization: Design space, opportunities, and challenges},
  author={Kim, Nam Wook and Joyner, Shakila Cherise and Riegelhuth, Amalia and Kim, Yeeun},
  booktitle={Computer graphics forum},
  volume={40},
  number={3},
  pages={173--188},
  year={2021},
  organization={Wiley Online Library}
}

@inproceedings{xie2023prompt,
  title={A prompt log analysis of text-to-image generation systems},
  author={Xie, Yutong and Pan, Zhaoying and Ma, Jinge and Jie, Luo and Mei, Qiaozhu},
  booktitle={Proceedings of the ACM Web Conference 2023},
  pages={3892--3902},
  year={2023}
}

@inproceedings{song2024real,
  title={A real-time chart explanation system for visually impaired individuals},
  author={Song, YooJeong and Jeong, SungHeon and Cho, WooJin and Lim, Soon-Bum and Park, Joo Hyun},
  booktitle={International Conference on Computers Helping People with Special Needs},
  pages={306--312},
  year={2024},
  organization={Springer}
}

@inproceedings{moured2024alt4blind,
  title={{Alt4Blind: a user interface to simplify charts alt-text creation}},
  author={Moured, Omar and Farooqui, Shahid Ali and M{\"u}ller, Karin and Fadaeijouybari, Sharifeh and Schwarz, Thorsten and Javed, Mohammed and Stiefelhagen, Rainer},
  booktitle={International Conference on Computers Helping People with Special Needs},
  pages={291--298},
  year={2024},
  organization={Springer}
}

@article{page2021prisma,
	author = {Page, Matthew J and McKenzie, Joanne E and Bossuyt, Patrick M and Boutron, Isabelle and Hoffmann, Tammy C and Mulrow, Cynthia D and Shamseer, Larissa and Tetzlaff, Jennifer M and Akl, Elie A and Brennan, Sue E and Chou, Roger and Glanville, Julie and Grimshaw, Jeremy M and Hr{\'o}bjartsson, Asbj{\o}rn and Lalu, Manoj M and Li, Tianjing and Loder, Elizabeth W and Mayo-Wilson, Evan and McDonald, Steve and McGuinness, Luke A and Stewart, Lesley A and Thomas, James and Tricco, Andrea C and Welch, Vivian A and Whiting, Penny and Moher, David},
	title = {{The PRISMA 2020 statement: an updated guideline for reporting systematic reviews}},
	volume = {372},
	elocation-id = {n71},
	year = {2021},
	doi = {10.1136/bmj.n71},
	publisher = {BMJ Publishing Group Ltd},
	URL = {https://www.bmj.com/content/372/bmj.n71},
	eprint = {https://www.bmj.com/content/372/bmj.n71.full.pdf},
	journal = {BMJ}
}

@article{whiting2016robis,
title = {{ROBIS: A new tool to assess risk of bias in systematic reviews was developed}},
journal = {Journal of Clinical Epidemiology},
volume = {69},
pages = {225-234},
year = {2016},
issn = {0895-4356},
doi = {10.1016/j.jclinepi.2015.06.005},
url = {https://www.sciencedirect.com/science/article/pii/S089543561500308X},
author = {Penny Whiting and Jelena Savović and Julian P.T. Higgins and Deborah M. Caldwell and Barnaby C. Reeves and Beverley Shea and Philippa Davies and Jos Kleijnen and Rachel Churchill},
}

@inproceedings{hsu2021scicap,
  title={{SciCap: Generating captions for scientific figures}},
  author={Hsu, Ting-Yao and Giles, C Lee and Huang, Ting-Hao},
  booktitle={Findings of the Association for Computational Linguistics: EMNLP 2021},
  pages={3258--3264},
  year={2021}
}

@article{yang2024scicap+,
  title={{Scicap+: A knowledge augmented dataset to study the challenges of scientific figure captioning}},
  author={Yang, Zhishen and Dabre, Raj and Tanaka, Hideki and Okazaki, Naoaki},
  journal={Journal of Natural Language Processing},
  volume={31},
  number={3},
  pages={1140--1165},
  year={2024},
  publisher={The Association for Natural Language Processing}
}

@article{melo2025impact,
  title={{The Impact of Artificial Intelligence on Inclusive Education: A Systematic Review}},
  author={Melo-L{\'o}pez, Ver{\'o}nica-Alexandra and Basantes-Andrade, Andrea and Gudi{\~n}o-Mej{\'\i}a, Carla-Bel{\'e}n and Hern{\'a}ndez-Mart{\'\i}nez, Evelyn},
  journal={Education Sciences},
  volume={15},
  number={5},
  pages={539},
  year={2025},
  publisher={MDPI}
}

@article{bernardi2016automatic,
  title={{Automatic description generation from images: A survey of models, datasets, and evaluation measures}},
  author={Bernardi, Raffaella and Cakici, Ruket and Elliott, Desmond and Erdem, Aykut and Erdem, Erkut and Ikizler-Cinbis, Nazli and Keller, Frank and Muscat, Adrian and Plank, Barbara},
  journal={Journal of Artificial Intelligence Research},
  volume={55},
  pages={409--442},
  year={2016}
}

@inproceedings{leotta2024evaluating,
  title={{Evaluating the Effectiveness of STEM Images Captioning}},
  author={Leotta, Maurizio and Ribaudo, Marina},
  booktitle={Proceedings of the 21st International Web for All Conference},
  pages={150--159},
  year={2024}
}

\end{document}